\documentclass[journal=jpcafh,manuscript=article]{achemso}
\setkeys{acs}{articletitle=true,doi=true}
\usepackage{amsmath,amssymb,bm}
\usepackage{booktabs,multirow}
\usepackage{graphicx}
\usepackage{tikz,pgfplots}
\pgfplotsset{compat=1.18}
\usetikzlibrary{arrows.meta}
\newif\ifitalian
\providecommand{\tr}[2]{\ifitalian#2\else#1\fi}

\definecolor{pblue}{RGB}{174,198,232}
\definecolor{pred}{RGB}{246,178,178}
\definecolor{pgreen}{RGB}{184,224,190}
\definecolor{pyellow}{RGB}{253,234,170}
\definecolor{ppurple}{RGB}{212,190,232}
\definecolor{pgray}{RGB}{238,238,242}
\definecolor{dblue}{RGB}{70,105,160}
\definecolor{dred}{RGB}{180,80,80}
\definecolor{dgreen}{RGB}{70,140,85}
\definecolor{dpurple}{RGB}{120,85,160}
\definecolor{dgray}{RGB}{110,110,120}

\tikzset{
  so/.style={draw=dgray,rounded corners=1.5pt,minimum width=7.5mm,minimum height=3.8mm,inner sep=0pt,font=\scriptsize},
  occ/.style={so,fill=pblue},
  emp/.style={so,fill=white},
  frc/.style={so,fill=pyellow},
  gai/.style={so,fill=pgreen},
  los/.style={so,fill=pred},
  pbox/.style={draw=dgray,rounded corners=3pt,align=center,inner sep=4pt,font=\footnotesize},
  parr/.style={-{Stealth[length=2mm]},line width=0.6pt,draw=dgray},
}

\newcommand{\FigConstructions}{%
\begin{tikzpicture}[font=\footnotesize]
\foreach \x/\sa/\sb/\k in {-0.75/occ/emp/1,0.75/emp/occ/2}{
  \node[occ] at (\x,2.40) {};
  \node[occ] at (\x,2.86) {};
  \node[\sa] at (\x,3.32) {};
  \node[\sb] at (\x,3.78) {};
  \node at (\x,1.98) {$\mathbf P_{\k},\ w_{\k}$};
}
\node[font=\scriptsize] at (0,3.32) {$i$};
\node[font=\scriptsize] at (0,3.78) {$a$};
\node[font=\footnotesize] at (0,4.3) {\tr{ensemble on common orbitals}{insieme su orbitali comuni}};
\node[occ] at (4.9,2.40) {};
\node[occ] at (4.9,2.86) {};
\node[frc] at (4.9,3.32) {$w_1$};
\node[frc] at (4.9,3.78) {$w_2$};
\node at (4.9,1.98) {$\bar{\mathbf P}=\sum_Iw_I\mathbf P_I$};
\node[pbox,fill=pgray] (ek) at (-4.9,3.09) {$E[\mathbf P_1],\ E[\mathbf P_2]$};
\draw[parr] (-1.35,3.09) -- node[above,font=\scriptsize]{\tr{energy of each}{energia di ciascun}} node[below,font=\scriptsize]{\tr{determinant}{determinante}} (ek.east);
\draw[parr] (1.35,3.09) -- node[above,font=\scriptsize]{\tr{average the}{media delle}} node[below,font=\scriptsize]{\tr{density matrices}{matrici densit\`a}} (4.25,3.09);
\node[pbox,fill=pgreen,minimum width=3.6cm] (sa) at (-4.9,0.55) {\textbf{SA-UHF}\\[1pt] $E_{\mathrm{SA}}=\sum_Iw_IE[\mathbf P_I]$};
\node[pbox,fill=pyellow,minimum width=3.6cm] (fon) at (4.9,0.55) {\textbf{FON-UHF}\\[1pt] $E_{\mathrm{FON}}=E[\bar{\mathbf P}]$};
\draw[parr] (ek.south) -- node[right,font=\scriptsize]{\tr{average}{media}} (sa.north);
\draw[parr] (4.9,1.72) -- node[right,font=\scriptsize]{\tr{energy of the average}{energia della media}} (fon.north);
\node[pbox,fill=ppurple] (g) at (0,0.55) {\tr{ghost interaction}{interazione fantasma}\\[1pt] $G=E_{\mathrm{FON}}-E_{\mathrm{SA}}$\\[1pt] $=-\tfrac12\sum_{I<J}w_Iw_JB(\Delta\mathbf P_{IJ},\Delta\mathbf P_{IJ})$};
\draw[{Stealth[length=2mm]}-{Stealth[length=2mm]},line width=0.6pt,draw=dpurple] (sa.east) -- (g.west);
\draw[{Stealth[length=2mm]}-{Stealth[length=2mm]},line width=0.6pt,draw=dpurple] (g.east) -- (fon.west);
\end{tikzpicture}}

\newcommand{\IndCol}[7]{%
  \node[occ] at (#1,0.9) {}; \node[occ] at (#1,1.35) {};
  \node[#2] at (#1,1.8) {}; \node[#3] at (#1,2.25) {};
  \node[font=\scriptsize,text=dgray] at (#1,2.75) {$\mathbf o=#7$};
  \node at (#1,0.38) {$#4$};
  \node[font=\scriptsize] at (#1,-0.12) {$#5$};
  \node[font=\scriptsize,text=dgray] at (#1,-0.55) {$N=#6$};}
\newcommand{\HCol}[6]{%
  \node[#2] at (#1,-2.25) {}; \node[#3] at (#1,-1.8) {};
  \node[font=\scriptsize,text=dgray] at (#1,-1.32) {$#6$};
  \node at (#1,-2.75) {#4};
  \node[font=\scriptsize] at (#1,-3.2) {$#5$};}
\newcommand{\FigIndependent}{%
\begin{tikzpicture}[font=\footnotesize]
\node[font=\footnotesize\bfseries] at (-1.55,2.75) {(a)};
\node[occ] at (0,0.9) {}; \node[occ] at (0,1.35) {};
\node[frc] at (0,1.8) {$n_p$}; \node[frc] at (0,2.25) {$n_q$};
\node[anchor=east,font=\scriptsize] at (-0.45,2.25) {$\varphi_q$};
\node[anchor=east,font=\scriptsize] at (-0.45,1.8) {$\varphi_p$};
\node[anchor=east,font=\scriptsize,align=right] at (-0.45,1.125) {\tr{integer}{intere}};
\node[font=\scriptsize,align=center] at (0,2.75) {FON-UHF};
\node at (0,0.38) {$\mathbf P(\mathbf n)$};
\node[font=\small] at (1.25,1.6) {$=$};
\node[font=\scriptsize,align=center] at (1.25,1.05) {\tr{average}{media}\\\tr{over}{su}};
\IndCol{2.6}{emp}{emp}{\mathbf P_0}{(1-n_p)(1-n_q)}{N_0}{(0,0)}
\IndCol{4.9}{occ}{emp}{\mathbf P(1,0)}{n_p(1-n_q)}{N_0+1}{(1,0)}
\IndCol{7.2}{emp}{occ}{\mathbf P(0,1)}{(1-n_p)n_q}{N_0+1}{(0,1)}
\IndCol{9.5}{occ}{occ}{\mathbf P(1,1)}{n_pn_q}{N_0+2}{(1,1)}
\draw[decorate,decoration={brace,amplitude=4pt},dgray] (1.9,3.05) -- (10.2,3.05)
  node[midway,above=4pt,font=\scriptsize,text=black] {\tr{$2^m=4$ determinants, each spin-orbital occupied independently with probability $n_p$}{$2^m=4$ determinanti, ogni spin-orbitale occupato indipendentemente con probabilit\`a $n_p$}};
\node[anchor=west,font=\scriptsize] at (10.4,0.38) {\tr{determinant}{determinante}};
\node[anchor=west,font=\scriptsize] at (10.4,-0.12) {\tr{weight}{peso} $W(\mathbf o)$};
\node[anchor=west,font=\scriptsize,text=dgray] at (10.4,-0.55) {\tr{electrons}{elettroni}};
\node[anchor=west,font=\scriptsize,text=dgray,align=left] at (10.4,2.0) {$o_q$\\[5pt]$o_p$};
\node[font=\footnotesize\bfseries] at (-1.55,-1.32) {(b)};
\node[frc] at (0,-2.25) {$\tfrac12$}; \node[frc] at (0,-1.8) {$\tfrac12$};
\node[anchor=east,font=\scriptsize] at (-0.45,-2.25) {$1s\alpha$};
\node[anchor=east,font=\scriptsize] at (-0.45,-1.8) {$1s\beta$};
\node at (0,-2.75) {H, FON};
\node[font=\small] at (1.25,-2.0) {$=$};
\HCol{2.6}{emp}{emp}{H$^+$}{\tfrac14}{(0,0)}
\HCol{4.9}{occ}{emp}{H ($\alpha$)}{\tfrac14}{(1,0)}
\HCol{7.2}{emp}{occ}{H ($\beta$)}{\tfrac14}{(0,1)}
\HCol{9.5}{occ}{occ}{H$^-$}{\tfrac14}{(1,1)}
\node[occ,minimum width=5mm,minimum height=3mm] at (2.4,-3.85) {}; \node[anchor=west,font=\scriptsize] at (2.75,-3.85) {\tr{occupied}{occupato}};
\node[emp,minimum width=5mm,minimum height=3mm] at (4.9,-3.85) {}; \node[anchor=west,font=\scriptsize] at (5.25,-3.85) {\tr{empty}{vuoto}};
\node[frc,minimum width=5mm,minimum height=3mm] at (7.2,-3.85) {}; \node[anchor=west,font=\scriptsize] at (7.55,-3.85) {\tr{fractional}{frazionario}};
\end{tikzpicture}}

\newcommand{\RankPanel}[4]{%
  \fill[pgray,rounded corners=4pt] (#1-1.75,-1.75) rectangle (#1+1.75,2.55);
  \node[font=\scriptsize\bfseries,align=center] at (#1,2.2) {#2};
  \node[font=\scriptsize] at (#1-0.5,-0.25) {$1-t$};
  \node[font=\scriptsize] at (#1+0.5,-0.25) {$t$};
  \node[font=\scriptsize,align=center] at (#1,-0.85) {#3};
  \node[font=\scriptsize,align=center] at (#1,-1.4) {#4};
}
\newcommand{\FigRank}{%
\begin{tikzpicture}[font=\scriptsize]
\RankPanel{0}{\tr{rank one}{rango uno}\\ $N\to N+1$}{$G=0$}{\tr{same theory}{stessa teoria}}
\node[emp] at (-0.5,0.9) {}; \node[gai] at (0.5,0.9) {};
\node[anchor=east] at (-1.0,0.9) {$p$};
\RankPanel{3.7}{\tr{rank two}{rango due}\\ $N\to N$}{$G=+t(1-t)A_{ia}$}{$G>0$}
\node[los] at (3.2,0.65) {}; \node[emp] at (4.2,0.65) {};
\node[emp] at (3.2,1.15) {}; \node[gai] at (4.2,1.15) {};
\node[anchor=east] at (2.7,0.65) {$i$}; \node[anchor=east] at (2.7,1.15) {$a$};
\RankPanel{7.4}{\tr{rank two}{rango due}\\ $N\to N+2$}{$G=-t(1-t)A_{ab}$}{$G<0$}
\node[emp] at (6.9,0.65) {}; \node[gai] at (7.9,0.65) {};
\node[emp] at (6.9,1.15) {}; \node[gai] at (7.9,1.15) {};
\node[anchor=east] at (6.4,0.65) {$a$}; \node[anchor=east] at (6.4,1.15) {$b$};
\RankPanel{11.1}{\tr{rank three}{rango tre}\\ $N\to N+1$}{\raisebox{0.6ex}{$G=t(1-t)\,\times$}\\ $(A_{ia}+A_{ib}-A_{ab})$}{\tr{sign not definite}{segno non definito}}
\node[los] at (10.6,0.45) {}; \node[emp] at (11.6,0.45) {};
\node[emp] at (10.6,0.95) {}; \node[gai] at (11.6,0.95) {};
\node[emp] at (10.6,1.45) {}; \node[gai] at (11.6,1.45) {};
\node[anchor=east] at (10.1,0.45) {$i$}; \node[anchor=east] at (10.1,0.95) {$a$}; \node[anchor=east] at (10.1,1.45) {$b$};
\node[gai,minimum width=5mm,minimum height=3mm] at (2.9,-2.2) {}; \node[anchor=west] at (3.25,-2.2) {\tr{gained}{guadagnato}};
\node[los,minimum width=5mm,minimum height=3mm] at (5.4,-2.2) {}; \node[anchor=west] at (5.75,-2.2) {\tr{lost}{perso}};
\node[emp,minimum width=5mm,minimum height=3mm] at (7.6,-2.2) {}; \node[anchor=west] at (7.95,-2.2) {\tr{empty}{vuoto}};
\end{tikzpicture}}

\usetikzlibrary{decorations.pathmorphing}
\tikzset{
  ghlev/.style={draw=dgray,line width=0.9pt},
  ghel/.style={-{Stealth[length=1.9mm,width=1.5mm]},line width=1.0pt,draw=dblue},
  ghfr/.style={-{Stealth[length=1.9mm,width=1.5mm]},line width=1.0pt,draw=dblue!70,densely dashed},
  ghreal/.style={decorate,decoration={snake,amplitude=1.1pt,segment length=4.2pt},line width=0.9pt,draw=dgreen},
  ghost/.style={decorate,decoration={snake,amplitude=1.1pt,segment length=4.2pt},line width=0.9pt,draw=dred},
}

\newcommand{\FigOxygen}{%
\begin{tikzpicture}[font=\footnotesize]
\foreach \x/\w/\ox/\oy in {0/{1-t}/0/0,2.6/{t(1-\lambda)}/1/0,5.2/{t\lambda}/0/1}{
  \fill[pgray,rounded corners=3pt] (\x-0.95,-0.45) rectangle (\x+0.95,1.0);
  \draw[ghlev] (\x-0.75,0.3) -- (\x-0.15,0.3); \draw[ghlev] (\x+0.15,0.3) -- (\x+0.75,0.3);
  \node[font=\scriptsize] at (\x-0.45,-0.25) {$\pi^*_x$}; \node[font=\scriptsize] at (\x+0.45,-0.25) {$\pi^*_y$};
  \draw[ghel] (\x-0.55,0.08)--(\x-0.55,0.52); \draw[ghel] (\x+0.35,0.08)--(\x+0.35,0.52);
  \ifnum\ox=1 \draw[ghel,draw=dgreen] (\x-0.35,0.52)--(\x-0.35,0.08);\fi
  \ifnum\oy=1 \draw[ghel,draw=dgreen] (\x+0.55,0.52)--(\x+0.55,0.08);\fi
  \node at (\x,1.25) {$\w$};
}
\node[font=\footnotesize\bfseries] at (2.6,1.75) {SA-UHF: \tr{three determinants}{tre determinanti}};
\fill[pyellow!55,rounded corners=3pt] (8.3,-0.45) rectangle (10.2,1.0);
\draw[ghlev] (8.5,0.3) -- (9.1,0.3); \draw[ghlev] (9.4,0.3) -- (10.0,0.3);
\node[font=\scriptsize] at (8.8,-0.25) {$\pi^*_x$}; \node[font=\scriptsize] at (9.7,-0.25) {$\pi^*_y$};
\draw[ghel] (8.7,0.08)--(8.7,0.52); \draw[ghel] (9.6,0.08)--(9.6,0.52);
\draw[ghfr,draw=dgreen!70] (8.9,0.52)--(8.9,0.08); \draw[ghfr,draw=dgreen!70] (9.8,0.52)--(9.8,0.08);
\node[font=\scriptsize] at (9.25,1.25) {$n_x=n_y=t/2$};
\node[font=\footnotesize\bfseries] at (9.25,1.75) {FON-UHF};
\end{tikzpicture}}

\newcommand{\Dlev}[2]{\draw[ghlev] (#1-0.27,#2) -- (#1+0.27,#2);}
\newcommand{\CoOct}[2]{%
  \begin{scope}[shift={(#1,#2)}]
    \foreach \a/\b in {0/0.62,0/-0.62,0.62/0,-0.62/0,0.33/0.2,-0.33/-0.2}{\draw[dgray,line width=0.6pt] (0,0) -- (\a,\b);}
    \draw[dgray!60,line width=0.4pt] (0.62,0) -- (0.33,0.2) -- (-0.62,0) -- (-0.33,-0.2) -- cycle;
    \foreach \a/\b in {0/0.62,0/-0.62,0.62/0,-0.62/0,0.33/0.2,-0.33/-0.2}{\fill[pblue!80!dblue,draw=dblue] (\a,\b) circle (0.085);}
    \fill[pred!70!dred,draw=dred] (0,0) circle (0.13);
    \node[font=\scriptsize] at (0,-0.92) {[Co(NH$_3$)$_6$]$^{3+/2+}$};
    \node[font=\tiny,text=dblue] at (0.48,0.72) {NH$_3$};
  \end{scope}}
\newcommand{\FigCobalt}{%
\begin{tikzpicture}[font=\footnotesize]
\CoOct{-2.2}{1.0}
\fill[pgray,rounded corners=3pt] (-1.05,-0.45) rectangle (1.75,2.55);
\node[align=center] at (0.35,3.0) {Co(III), $t_{2g}^6$\\ \tr{low spin}{basso spin} ($1-t$)};
\foreach \x in {-0.15,0.5,1.15}{\Dlev{\x}{0.3}}
\foreach \x in {0.17,0.83}{\Dlev{\x}{1.75}}
\node[font=\scriptsize,anchor=east] at (-0.5,0.3) {$t_{2g}$};
\node[font=\scriptsize,anchor=east] at (-0.1,1.75) {$e_g$};
\foreach \x in {-0.15,0.5,1.15}{\draw[ghel] (\x-0.08,0.08)--(\x-0.08,0.52); \draw[ghel] (\x+0.08,0.52)--(\x+0.08,0.08);}
\fill[pgray,rounded corners=3pt] (3.05,-0.45) rectangle (5.85,2.55);
\node[align=center] at (4.45,3.0) {Co(II), $t_{2g}^5e_g^2$\\ \tr{high spin}{alto spin} ($t$)};
\foreach \x in {3.95,4.6,5.25}{\Dlev{\x}{0.3}}
\foreach \x in {4.27,4.93}{\Dlev{\x}{1.75}}
\node[font=\scriptsize,anchor=east] at (3.6,0.3) {$t_{2g}$};
\node[font=\scriptsize,anchor=east] at (4.0,1.75) {$e_g$};
\foreach \x in {3.95,4.6}{\draw[ghel] (\x-0.08,0.08)--(\x-0.08,0.52); \draw[ghel] (\x+0.08,0.52)--(\x+0.08,0.08);}
\draw[ghel] (5.17,0.08)--(5.17,0.52);
\draw[dred,line width=1pt] (5.24,0.12)--(5.42,0.48); \draw[dred,line width=1pt] (5.42,0.12)--(5.24,0.48);
\draw[ghel,draw=dgreen] (4.27,1.53)--(4.27,1.97); \draw[ghel,draw=dgreen] (4.93,1.53)--(4.93,1.97);
\draw[parr] (1.9,1.05) -- node[above,font=\scriptsize]{$+e^-$} (2.9,1.05);
\node[anchor=west,align=left] at (6.3,2.1) {\textcolor{dgreen}{\tr{gained}{guadagnati}: $e_g\alpha$, $e_g'\alpha$} ($m_+=2$)};
\node[anchor=west,align=left] at (6.3,1.45) {\textcolor{dred}{\tr{lost}{perso}: $t_{2g}\beta$} ($m_-=1$)};
\node[anchor=west,align=left] at (6.3,0.8) {\tr{rank}{rango} $R=3$, \ $N_2-N_1=1$};
\node[anchor=west,align=left,font=\scriptsize] at (6.3,0.05) {$G=\tfrac14\big[\underbrace{A_{t_{2g}\beta,e_g\alpha}+A_{t_{2g}\beta,e_g'\alpha}}_{\text{\tr{spurious}{spuria}: }J+J}-\underbrace{A_{e_g\alpha,e_g'\alpha}}_{\text{\tr{missing}{mancante}: }J-K}\big]$};
\end{tikzpicture}}

\newcommand{\CoBox}[5]{%
  \fill[pgray,rounded corners=3pt] (#1-0.95,#2-0.3) rectangle (#1+0.95,#2+1.5);
  \foreach \k/\dx in {1/-0.6,2/0,3/0.6}{%
    \draw[ghlev] (#1+\dx-0.24,#2+0.2) -- (#1+\dx+0.24,#2+0.2);
    \draw[ghel] (#1+\dx-0.08,#2+0.02) -- (#1+\dx-0.08,#2+0.42);
    \ifnum\k=#3
      \draw[dred,line width=0.9pt] (#1+\dx+0.01,#2+0.08) -- (#1+\dx+0.17,#2+0.34);
      \draw[dred,line width=0.9pt] (#1+\dx+0.17,#2+0.08) -- (#1+\dx+0.01,#2+0.34);
    \else
      \draw[ghel] (#1+\dx+0.08,#2+0.42) -- (#1+\dx+0.08,#2+0.02);
    \fi}
  \foreach \dx in {-0.3,0.3}{%
    \draw[ghlev] (#1+\dx-0.24,#2+1.05) -- (#1+\dx+0.24,#2+1.05);
    \ifnum#4=1 \draw[ghel,draw=dgreen] (#1+\dx,#2+0.85) -- (#1+\dx,#2+1.27);\fi}
  \node[font=\scriptsize,align=center] at (#1,#2-0.62) {#5};}
\newcommand{\FigCoSym}{%
\begin{tikzpicture}[font=\footnotesize]
\node[font=\footnotesize\bfseries,anchor=west] at (-1.0,2.15) {\tr{(a) localized hole}{(a) lacuna localizzata}};
\CoBox{0}{0}{0}{0}{Co(III)\\ $1-t$}
\CoBox{2.6}{0}{1}{1}{Co(II)$[1]$\\ $t$}
\draw[parr,draw=dblue,<->] (0.98,0.6) -- node[above,font=\scriptsize]{\tr{rank 3}{rango 3}} (1.62,0.6);
\node[font=\footnotesize\bfseries,anchor=west] at (-1.0,-1.45) {\tr{(b) symmetric hole}{(b) lacuna simmetrica}};
\CoBox{0}{-4.2}{0}{0}{Co(III)\\ $1-t$}
\CoBox{2.6}{-4.2}{1}{1}{Co(II)$[1]$\\ $t/3$}
\CoBox{5.0}{-4.2}{2}{1}{Co(II)$[2]$\\ $t/3$}
\CoBox{7.4}{-4.2}{3}{1}{Co(II)$[3]$\\ $t/3$}
\draw[parr,draw=dblue,<->] (0.98,-3.6) -- node[above,font=\scriptsize]{\tr{rank 3}{rango 3}} (1.62,-3.6);
\draw[ghost] (2.6,-2.65) to[out=35,in=145] node[above,font=\scriptsize,text=dred]{\tr{rank 2}{rango 2}} (5.0,-2.65);
\draw[ghost] (5.0,-2.65) to[out=35,in=145] node[above,font=\scriptsize,text=dred]{\tr{rank 2}{rango 2}} (7.4,-2.65);
\draw[ghost] (2.6,-2.65) to[out=60,in=120] node[above,font=\scriptsize,text=dred]{\tr{rank 2}{rango 2}} (7.4,-2.65);
\node[anchor=west,align=left,font=\scriptsize] at (4.3,0.6) {\tr{gained (green): $e_g\alpha$, $e_g'\alpha$\\ lost (red cross): $t_{2g}\beta$\\[3pt] FON-UHF on (a): 3 fractional spin-orbitals, 8 determinants\\ FON-UHF on (b): 5 fractional spin-orbitals, 32 determinants}{guadagnati (verde): $e_g\alpha$, $e_g'\alpha$\\ perso (croce rossa): $t_{2g}\beta$\\[3pt] FON-UHF su (a): 3 spin-orbitali frazionari, 8 determinanti\\ FON-UHF su (b): 5 spin-orbitali frazionari, 32 determinanti}};
\end{tikzpicture}}

\newcommand{\FigBenzene}{%
\begin{tikzpicture}[font=\footnotesize]
\begin{scope}[shift={(-1.9,0.25)}]
  \foreach \k in {0,...,5}{\coordinate (v\k) at (60*\k:0.55);}
  \draw[dgray,line width=0.8pt] (v0)--(v1)--(v2)--(v3)--(v4)--(v5)--cycle;
  \draw[dgray,line width=0.5pt] (0,0) circle (0.3);
  \foreach \k in {0,...,5}{\draw[dgray!70,line width=0.5pt] (v\k) -- (60*\k:0.82);}
  \node[font=\scriptsize] at (0,-1.05) {C$_6$H$_6$, $D_{6h}$};
\end{scope}
\foreach \x/\w/\ox/\oy in {0/{1-t}/0/0,2.6/{t(1-\lambda)}/1/0,5.2/{t\lambda}/0/1}{
  \fill[pgray,rounded corners=3pt] (\x-0.95,-0.45) rectangle (\x+0.95,1.0);
  \draw[ghlev] (\x-0.75,0.3) -- (\x-0.15,0.3); \draw[ghlev] (\x+0.15,0.3) -- (\x+0.75,0.3);
  \node[font=\scriptsize] at (\x-0.45,-0.25) {$e_{2u,x}$}; \node[font=\scriptsize] at (\x+0.45,-0.25) {$e_{2u,y}$};
  \ifnum\ox=1 \draw[ghel,draw=dgreen] (\x-0.45,0.08)--(\x-0.45,0.52);\fi
  \ifnum\oy=1 \draw[ghel,draw=dgreen] (\x+0.45,0.08)--(\x+0.45,0.52);\fi
  \node at (\x,1.25) {$\w$};
}
\node[font=\footnotesize\bfseries] at (2.6,1.75) {SA-UHF: \tr{three determinants}{tre determinanti}};
\fill[pyellow!55,rounded corners=3pt] (8.3,-0.45) rectangle (10.2,1.0);
\draw[ghlev] (8.5,0.3) -- (9.1,0.3); \draw[ghlev] (9.4,0.3) -- (10.0,0.3);
\node[font=\scriptsize] at (8.8,-0.25) {$e_{2u,x}$}; \node[font=\scriptsize] at (9.7,-0.25) {$e_{2u,y}$};
\draw[ghfr,draw=dgreen!70] (8.8,0.08)--(8.8,0.52); \draw[ghfr,draw=dgreen!70] (9.7,0.08)--(9.7,0.52);
\draw[ghost] (8.85,0.62) to[out=60,in=120] (9.65,0.62);
\node[font=\scriptsize] at (9.25,1.25) {$n_x=n_y=t/2$};
\node[font=\footnotesize\bfseries] at (9.25,1.75) {FON-UHF};
\node[font=\scriptsize,text=dgray,anchor=north] at (4.7,-0.6) {\tr{cation: the same with a hole shared between the $e_{1g}$ components}{catione: lo stesso con una lacuna condivisa tra le componenti $e_{1g}$}};
\end{tikzpicture}}

\newcommand{\Eth}[3]{%
  \fill[#3,opacity=0.9] (#1,#2) ellipse (0.62 and 0.17);
  \draw[dgray,line width=1.1pt] (#1-0.34,#2) -- (#1+0.34,#2);
  \foreach \s in {-1,1}{\draw[dgray,line width=0.6pt] (#1+\s*0.34,#2) -- (#1+\s*0.62,#2+0.12); \draw[dgray,line width=0.6pt] (#1+\s*0.34,#2) -- (#1+\s*0.62,#2-0.12);}
  \fill[dgray] (#1-0.34,#2) circle (0.05); \fill[dgray] (#1+0.34,#2) circle (0.05);}
\newcommand{\FigDimer}{%
\begin{tikzpicture}[font=\footnotesize]
\node[font=\footnotesize\bfseries] at (2.0,2.35) {SA-UHF: \tr{the hole is on $A$ or on $B$}{la lacuna \`e su $A$ oppure su $B$}};
\foreach \x/\ha/\hb in {0.8/1/0,3.2/0/1}{
  \fill[pgray,rounded corners=3pt] (\x-0.95,-0.15) rectangle (\x+0.95,1.95);
  \ifnum\ha=1 \def\ca{pred}\else\def\ca{pblue!60}\fi
  \ifnum\hb=1 \def\cb{pred}\else\def\cb{pblue!60}\fi
  \Eth{\x}{0.35}{\ca}
  \Eth{\x}{1.45}{\cb}
  \node[font=\scriptsize,anchor=west] at (\x+0.62,0.35) {$A$};
  \node[font=\scriptsize,anchor=west] at (\x+0.62,1.45) {$B$};
}
\node at (0.8,-0.4) {$\tfrac12$}; \node at (3.2,-0.4) {$\tfrac12$}; \node at (2.0,0.9) {$+$};
\draw[{Stealth[length=1.4mm]}-{Stealth[length=1.4mm]},dgray,line width=0.5pt] (-0.35,0.35) -- node[left,font=\scriptsize]{$R$} (-0.35,1.45);
\node[font=\footnotesize\bfseries] at (6.3,2.35) {FON-UHF};
\fill[pyellow!55,rounded corners=3pt] (5.35,-0.15) rectangle (7.25,1.95);
\Eth{6.3}{0.35}{pred!55} \Eth{6.3}{1.45}{pred!55}
\node[font=\scriptsize,anchor=west] at (6.92,0.35) {$\tfrac12$}; \node[font=\scriptsize,anchor=west] at (6.92,1.45) {$\tfrac12$};
\draw[ghost] (6.3,0.55) -- (6.3,1.25);
\node[anchor=west,align=left,text=dred] at (7.6,0.9) {\tr{two half holes that are}{due mezze lacune che}\\ \tr{never present together:}{non sono mai presenti insieme:}\\ $G=\tfrac14A_{AB}\to\dfrac{1}{4R}$};
\end{tikzpicture}}

\newcommand{\GhMini}[4]{%
  \fill[pgray,rounded corners=2pt] (#1-0.42,#2-0.42) rectangle (#1+0.42,#2+0.42);
  \draw[ghlev] (#1-0.3,#2) -- (#1+0.3,#2);
  \ifnum#3=1 \draw[ghel] (#1-0.1,#2-0.24) -- (#1-0.1,#2+0.26);\fi
  \ifnum#4=1 \draw[ghel] (#1+0.1,#2+0.26) -- (#1+0.1,#2-0.24);\fi
}

\newcommand{\TocGhost}[3]{%
  \begin{scope}[shift={(#1,#2)},scale=#3]
    \fill[black,opacity=0.12] (0.03,-0.33) ellipse (0.24 and 0.05);
    \draw[dred!80!black,line width=0.5pt,fill=white,fill opacity=0.95,rounded corners=0.6pt]
      (-0.22,-0.24) -- (-0.22,0.05) arc (180:0:0.22) -- (0.22,-0.24)
      -- (0.147,-0.17) -- (0.073,-0.24) -- (0,-0.17) -- (-0.073,-0.24) -- (-0.147,-0.17) -- cycle;
    \fill[dgray] (-0.075,0.07) ellipse (0.032 and 0.048);
    \fill[dgray] (0.075,0.07) ellipse (0.032 and 0.048);
    \fill[white] (-0.065,0.085) circle (0.011); \fill[white] (0.085,0.085) circle (0.011);
    \draw[dgray,line width=0.35pt] (0,-0.05) ellipse (0.028 and 0.036);
    \fill[pred!80] (-0.13,-0.03) ellipse (0.03 and 0.017); \fill[pred!80] (0.13,-0.03) ellipse (0.03 and 0.017);
  \end{scope}}
\newcommand{\TocLevels}[2]{\foreach \yy in {0,0.33,0.66}{\draw[dgray,line width=0.6pt] (#1-0.22,#2+\yy) -- (#1+0.22,#2+\yy);}}
\newcommand{\TocUp}[3]{\draw[-{Stealth[length=1.3mm,width=1.1mm]},line width=0.75pt,#1] (#2,#3-0.12) -- (#2,#3+0.14);}
\newcommand{\FigTOC}{%
\begingroup\def\baselinestretch{1}\resizebox{3.25in}{!}{\begin{tikzpicture}[x=1.5cm,y=1.5cm,font=\scriptsize]
\useasboundingbox (0,0) rectangle (8.25,4.43);
\shade[top color=white,bottom color=pblue!45,rounded corners=4pt] (0,0) rectangle (8.25,4.43);
\node[font=\small\bfseries,text=dgray!80!black] at (4.125,4.16) {Is the energy of the average the average of energies?};
\begin{scope}[shift={(0,-0.32)}]
\foreach \x/\lab in {0.62/{$\mathbf P_I,\,1{-}t$},1.88/{$\mathbf P_J,\,t$}}{
  \fill[white,rounded corners=3pt,draw=dgray!50] (\x-0.4,0.92) rectangle (\x+0.4,2.02);
  \TocLevels{\x}{1.12}
  \node[font=\scriptsize] at (\x,0.8) {\lab};}
\node[font=\scriptsize,anchor=east,text=dgray,inner sep=1pt] at (0.21,1.12) {$i$};
\node[font=\scriptsize,anchor=east,text=dgray,inner sep=1pt] at (0.21,1.45) {$a$};
\node[font=\scriptsize,anchor=east,text=dgray,inner sep=1pt] at (0.21,1.78) {$b$};
\TocUp{dblue}{0.62}{1.12}
\TocUp{dgreen}{1.88}{1.45}\TocUp{dgreen}{1.88}{1.78}
\node[font=\normalsize] at (1.25,1.47) {$+$};
\fill[pyellow,rounded corners=3pt,draw=orange!55!dgray] (0.85,2.62) rectangle (1.65,3.72);
\TocLevels{1.25}{2.82}
\foreach \yy/\w in {2.82/{1{-}t},3.15/t,3.48/t}{
  \draw[-{Stealth[length=1.2mm,width=1mm]},line width=0.7pt,dred,densely dashed] (1.25,\yy-0.12) -- (1.25,\yy+0.14);
  \node[font=\scriptsize,anchor=west,inner sep=0.5pt] at (1.48,\yy+0.07) {$\w$};}
\node[font=\scriptsize,anchor=east] at (0.82,3.17) {$\bar{\mathbf P}$};
\draw[-{Stealth[length=1.5mm]},line width=0.6pt,dgray] (1.25,2.1) -- node[left,font=\scriptsize,text=dgray,inner sep=1pt]{average} (1.25,2.55);
\draw[-{Stealth[length=2mm]},line width=1.5pt,dred,rounded corners=4pt] (1.7,3.17) -- (3.72,3.17);
\node[font=\scriptsize,text=dred,anchor=south,align=center,inner sep=1pt] at (2.7,3.2) {energy of\\the average};
\draw[-{Stealth[length=2mm]},line width=1.5pt,dblue,rounded corners=4pt] (2.3,1.47) -- (3.72,1.47);
\node[font=\scriptsize,text=dblue,anchor=south,align=center,inner sep=1pt] at (3.0,1.5) {average of\\energies};
\begin{scope}[shift={(3.95,0.95)}]
  \fill[white,opacity=0.75,rounded corners=3pt] (-0.15,-0.25) rectangle (4.2,2.95);
  \draw[-{Stealth[length=1.5mm]},dgray,line width=0.5pt] (0,0) -- (4.05,0);
  \node[font=\scriptsize,anchor=north east,inner sep=1pt] at (4.05,0) {$t$};
  \draw[-{Stealth[length=1.5mm]},dgray,line width=0.5pt] (0,0) -- (0,2.85) node[below right,font=\scriptsize,inner sep=1pt] {$E$};
  \node[font=\scriptsize,below,inner sep=1pt] at (0.15,0) {$0$}; \node[font=\scriptsize,below,inner sep=1pt] at (3.65,0) {$1$};
  \fill[pred!55,domain=0:1,samples=40] plot ({0.15+3.5*\x},{0.35+1.0*\x}) -- plot[domain=1:0] ({0.15+3.5*\x},{0.35+1.0*\x+5.6*\x*(1-\x)});
  \draw[dred,line width=1.6pt,domain=0:1,samples=60] plot ({0.15+3.5*\x},{0.35+1.0*\x+5.6*\x*(1-\x)});
  \draw[dblue,line width=1.9pt] (0.15,0.35) -- (3.65,1.35);
  \fill[dblue] (0.15,0.35) circle (1.5pt); \fill[dblue] (3.65,1.35) circle (1.5pt);
  \node[font=\scriptsize,text=dred,anchor=south,inner sep=1pt] at (1.9,2.35) {$E_{\mathrm{FON}}=E[\bar{\mathbf P}]$};
  \node[font=\scriptsize,text=dblue,anchor=north,inner sep=1pt,rotate=16] at (2.25,0.92) {$E_{\mathrm{SA}}=\sum_Iw_IE[\mathbf P_I]$};
  \TocGhost{1.35}{1.55}{1.25}
  \node[font=\scriptsize,text=dred!85!black,anchor=west,align=left,inner sep=1pt] at (1.72,1.72) {\textbf{ghost}\\[-1pt]$G=E_{\mathrm{FON}}-E_{\mathrm{SA}}$};
\end{scope}
\end{scope}
\end{tikzpicture}}\endgroup}

\usetikzlibrary{decorations.pathreplacing}
\newcommand{\HidArc}[5]{%
  \ifx#5l\draw[#1] (#2-0.42,#3) to[out=180,in=180,looseness=1.6] (#2-0.42,#4);
  \else\draw[#1] (#2+0.42,#3) to[out=0,in=0,looseness=1.6] (#2+0.42,#4);\fi}
\newcommand{\FigHidden}{%
\begin{tikzpicture}[font=\footnotesize,
  chg/.style={{Stealth[length=1.6mm]}-{Stealth[length=1.6mm]},line width=1pt,draw=dred},
  inv/.style={{Stealth[length=1.6mm]}-{Stealth[length=1.6mm]},line width=0.7pt,draw=dgray,densely dashed}]
\foreach \x/\si/\sa/\sb/\lab in {0/occ/emp/emp/{$\mathbf P_I,\ \tfrac12$},2.0/emp/occ/occ/{$\mathbf P_J,\ \tfrac12$}}{
  \node[\si] at (\x,0) {}; \node[\sa] at (\x,0.7) {}; \node[\sb] at (\x,1.4) {};
  \node at (\x,2.05) {\lab};
  \HidArc{chg}{\x}{0}{0.7}{r}
  \HidArc{inv}{\x}{0.7}{1.4}{l}}
\node[frc] at (4.9,0) {$\tfrac12$}; \node[frc] at (4.9,0.7) {$\tfrac12$}; \node[frc] at (4.9,1.4) {$\tfrac12$};
\node at (4.9,2.05) {$\bar{\mathbf P}$};
\HidArc{inv}{4.9}{0}{0.7}{r}
\HidArc{inv}{4.9}{0.7}{1.4}{l}
\foreach \y/\o in {0/i,0.7/a,1.4/b}{\node[font=\scriptsize,anchor=east] at (-0.85,\y) {$\o$};}
\draw[parr] (3.0,0.7) -- node[above,font=\scriptsize]{\tr{average}{media}} (4.15,0.7);
\draw[dgray,decorate,decoration={brace,mirror,amplitude=4pt}] (-0.6,-0.45) -- (2.6,-0.45);
\node[align=center,font=\scriptsize] at (1.0,-1.05) {\textbf{SA-UHF}\\ \tr{sees each determinant}{vede ogni determinante}};
\draw[dgray,decorate,decoration={brace,mirror,amplitude=4pt}] (4.3,-0.45) -- (5.5,-0.45);
\node[align=center,font=\scriptsize] at (4.9,-1.05) {\textbf{FON-UHF}\\ \tr{sees only}{vede solo} $\bar{\mathbf P}$};
\begin{scope}[shift={(6.9,0)}]
\draw[chg] (0,1.55) -- (0.6,1.55);
\node[anchor=west,align=left,font=\scriptsize] at (0.75,1.55) {\tr{rotation changes the matrix}{la rotazione cambia la matrice}};
\draw[inv] (0,1.05) -- (0.6,1.05);
\node[anchor=west,align=left,font=\scriptsize] at (0.75,1.05) {\tr{rotation leaves it unchanged}{la rotazione la lascia invariata}};
\node[anchor=west,align=left,font=\scriptsize,text width=5.6cm] at (0,0.35) {$(a,b)$: \tr{same occupation class,\\ redundant for both constructions}{stessa classe di occupazione,\\ ridondante per entrambe le costruzioni}};
\node[anchor=west,align=left,font=\scriptsize,text width=5.6cm,text=dred!85!black] at (0,-0.75) {$(i,a)$, $(i,b)$: \textbf{\tr{hidden pairs}{coppie nascoste}}\\ $\bar n_i=\bar n_a$ \tr{but different patterns}{ma schemi diversi}\\ $g_{ia}=\tfrac12[(ia|bb)-(ib|ba)]$, \tr{in general}{in generale} $\neq0$};
\end{scope}
\end{tikzpicture}}

\usepackage{adjustbox}
\usepackage[version=4]{mhchem}
\usepackage[labelfont=bf,textfont=normalfont,labelsep=colon,font=small]{caption}
\usepackage{hyperref}
\makeatletter
\newcommand{\doi}{\begingroup\catcode`\_=12\catcode`\%=12\catcode`\&=12\catcode`\~=12\catcode`\#=12\doi@link}
\newcommand{\doi@link}[1]{\href{https://doi.org/#1}{\texttt{#1}}\endgroup}
\makeatother
\hypersetup{colorlinks=true,linkcolor=blue,citecolor=blue,urlcolor=blue}
\SectionNumbersOn

\newcommand{\Pb}{\bar{\mathbf{P}}}
\newcommand{\Fb}{\bar{\mathbf{F}}}
\newcommand{\Pt}{\widetilde{\mathbf{P}}}
\newcommand{\dP}{\Delta\mathbf{P}}
\newcommand{\kets}[1]{|\varphi_{#1}\rangle}
\newcommand{\proj}[1]{|\varphi_{#1}\rangle\langle\varphi_{#1}|}
\newcommand{\Esa}{E_{\mathrm{SA}}}
\newcommand{\Efon}{E_{\mathrm{FON}}}
\newcommand{\dEgh}{\Delta E_{\mathrm{gh}}}
\newcommand{\Cov}{\mathrm{Cov}}

\author{Gabriele Fabbro}
\email{gabriele.fabbro@ntnu.no}
\author{Ida-Marie H{\o}yvik}
\affiliation{Department of Chemistry, Norwegian University of Science and Technology, NO-7491 Trondheim, Norway}

\title{The Ensemble Ghost Interaction in Fractional-Occupation Hartree--Fock: Occupation Covariance, Pair Rank, and Prescribed Ensembles of Determinants}

\begin{document}

\begin{abstract}
Fractional Occupation Number unrestricted Hartree--Fock (FON-UHF) evaluates the UHF functional on a density matrix with non-integer occupations. Sector-Averaged UHF (SA-UHF) averages the energies of a prescribed ensemble of determinants, possibly from different particle-number sectors, on common orbitals. The two differ by the ensemble ghost interaction $G$. We show that, for two determinants, the numbers of spin-orbitals gained and lost between them classify $G$ before any calculation. $G$ vanishes on every orbital set at pair rank one, has a fixed sign at pair rank two and whenever two or more spin-orbitals are only gained or only lost, and otherwise has an orbital-dependent sign. For larger ensembles, $G$ vanishes on every orbital set when all pair covariances of the occupations vanish, and in the complete-basis limit only then. The same pairs add a covariance term to the orbital gradient and give the error of the chemical potential estimated from the ensemble Fock matrix. We implemented both methods in the eT program. The calculations give $G=0.24$~Hartree at half transfer for the hexaamminecobalt(III/II) couple, 78--84\% of the FON-UHF penalty for a fraction shared between the degenerate frontier orbitals of benzene, and the limit $1/(4R)$ for a hole shared between two stacked ethylene molecules.
\end{abstract}

\begin{center}
\textbf{TOC Graphic}\par\medskip
\fbox{\includegraphics[width=3.25in]{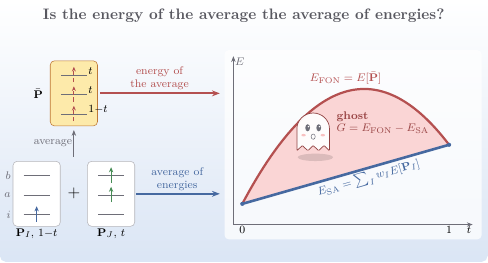}}
\end{center}

\clearpage
\section{Introduction}

Adding a fraction of an electron to a molecule has a precise meaning only for an ensemble of states with different electron numbers. Perdew, Parr, Levy and Balduz (PPLB) showed that, if the exact energies are convex in the electron number, the lowest-energy ensemble combines the two adjacent integers and is linear in the fractional weight.\cite{Perdew1982} Deviations from this piecewise linearity diagnose delocalization and static-correlation errors. Semilocal functionals are convex between integers and Hartree--Fock (HF) is concave,\cite{Vydrov2007,MoriSanchez2008PRL,Cohen2008Science,LiYang2017,Peach2015} with analogous conditions for fractional spin and the ``flat plane''.\cite{Yang2000,Cohen2008FracSpin,MoriSanchez2009} At frozen orbitals the HF energy is linear in a single fractional occupation, and orbital relaxation can only lower the curvature.\cite{LiYang2017}

Given an ensemble with a fractional electron number, its HF energy can be constructed in two ways. The older construction evaluates the energy of a density matrix with fractional occupations. The hyper-Hartree--Fock method of Slater and co-workers extended the average energy of a configuration to non-integral subshell occupations\cite{Slater1969} and led to the transition-state concept.\cite{Slater1972} Abdulnur and co-workers defined open-shell functionals as ensemble averages.\cite{Abdulnur1972} They distinguished the average over a product density operator from its projection onto a fixed electron number. In the product average, pair occupations factorize, a single Fock operator results and orbital energies are occupation derivatives; after the projection, none of these holds. J{\o}rgensen and {\"O}hrn called the two schemes grand-canonical (GC) and canonical HF and carried them into time-dependent and propagator theory.\cite{Jorgensen1972,JorgensenOhrn1973PRA,JorgensenOhrn1973CPL,LinderbergOhrn1973} Zerner observed that in GC HF ``fractional electrons are allowed to interact with one another'', so that its energy lies above the configuration average.\cite{Zerner1989} Fractional occupations are also used as a convergence device,\cite{RabuckScuseria1999} in thermal HF,\cite{Mermin1963,GuHirata2024} in thermally-assisted-occupation DFT and the fractional-occupation density diagnostic,\cite{Chai2012,GrimmeHansen2015} and in transition-operator references for ionization.\cite{Brandi1980,Baiardi2017} Because the spin-orbital occupations of the product ensemble are independent, its energy is the HF functional of its fractionally occupied density matrix, a finite-temperature form of Wick's theorem.\cite{Gaudin1960} The same ensemble-averaged reference underlies the extension of many-body perturbation theory and random-phase approximations to fractional charges and spins.\cite{Yang2013} We call the unrestricted version Fractional Occupation Number UHF (FON-UHF).

\begin{figure}
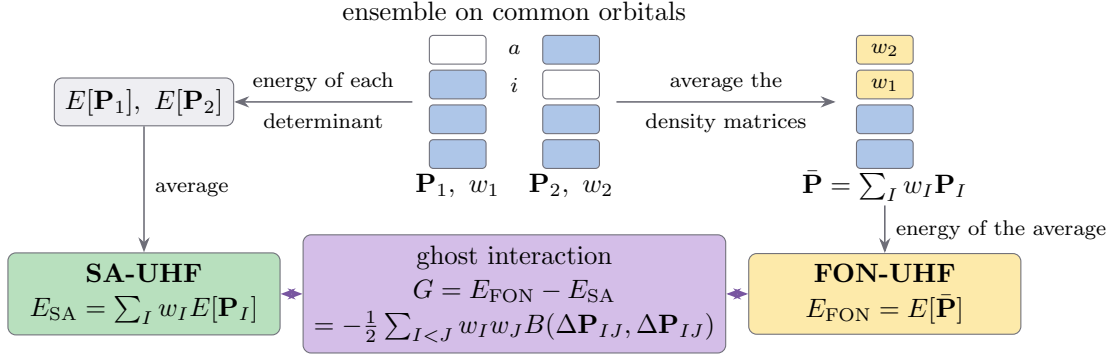

\centering
\FigConstructions
\caption{The two constructions compared in this work, for an ensemble of two determinants that differ in the spin-orbitals $i$ and $a$. SA-UHF averages the energies of the determinants; FON-UHF evaluates the UHF functional on the averaged density matrix, whose occupations are fractional. The two differ by the ensemble ghost interaction $G$, Eq.~\ref{eq:Gpair}, which can have either sign (Figure~\ref{fig:rank}).}
\label{fig:scheme}
\end{figure}

The second construction averages the \emph{energies} of determinants. For ensembles of ground and excited states,\cite{Theophilou1979,GOK1988} Gidopoulos and co-workers identified the ``ghost interaction'' that appears when the Hartree and exchange energies are evaluated on the ensemble density. They noted that ensemble HF, an average of determinant energies on common orbitals, is free of it.\cite{Gidopoulos2002PRL,Gidopoulos2002PhysB} Pastorczak and Pernal wrote the correction for ensembles of many states as a sum over pairs of states,\cite{PastorczakPernal2014,Pastorczak2013} and Alam and co-workers derived it in range-separated ensemble DFT.\cite{Alam2016} Gould and co-workers built ghost-free Hartree and exchange energies from pair coefficients of the ensemble;\cite{GouldPittalis2017,GouldKronik2021,GouldKronikPittalis2026} the contrast with state-averaged HF is reviewed in ref~\citenum{Cernatic2022}. Across particle-number sectors, the linear exact exchange (LEXX) of Gould and Dobson averages the energies of determinants from adjacent electron numbers on common orbitals (from $N$ to $N+2$ for a fractional $s$ orbital). It identifies the ghost terms of a fractional $s$ orbital or $p$ shell through pair occupations.\cite{GouldDobson2013} The ensemble generalization of Kraisler and Kronik\cite{Kraisler2013,Kraisler2014} and $N$-centred ensembles\cite{SenjeanFromager2018} average Hartree and exchange--correlation energies over adjacent sectors; for a single fractional spin-orbital with exact exchange, the Hartree and exchange corrections cancel.\cite{Kraisler2013} We call the HF version of this construction Sector-Averaged UHF (SA-UHF): UHF determinants of integer-electron sectors on one common orbital set, with prescribed weights (Figure~\ref{fig:scheme}).

The two constructions coincide for a single fractional spin-orbital and in general differ when there are several. A minimum of FON-UHF over the occupations has at most one fractional occupation for each constrained particle number.\cite{LiLuYang2017,BLLS1994,Lieb1981} The difference therefore matters only for \emph{prescribed} ensembles, whose sectors, spin states and occupation patterns are fixed in advance. Examples are a fraction shared by symmetry among degenerate orbitals\cite{JahnTeller1937} or equivalent centres, as in mixed-valence compounds\cite{RobinDay1968} and broken-symmetry descriptions of transition-metal dimers;\cite{Noodleman1981} the two members of a redox couple that differ in spin state;\cite{Hammershoi1984} and a non-adjacent sector pair. Fermi smearing and fictitious electronic temperatures also give many fractional occupations at once.\cite{RabuckScuseria1999,Chai2012,GrimmeHansen2015}

Most ingredients of the comparison are known and are restated here only in the form needed. They include the pairwise and covariance forms of the ghost interaction,\cite{Abdulnur1972,PastorczakPernal2014,Alam2016,GouldDobson2013} the independence of the spin-orbital occupations in FON-UHF,\cite{Abdulnur1972,Gaudin1960} and the dependence of the averaged energy on the ensemble and not only on its density matrix.\cite{GouldPittalis2017,GouldKronik2021,GouldKronikPittalis2026} They also include the stationarity conditions of state-averaged methods, which cannot in general be reduced to the diagonalization of one Fock matrix,\cite{Abdulnur1972,JorgensenOhrn1973PRA,Gidopoulos2002PhysB,Cernatic2022} and the non-redundancy of rotations between equally occupied orbitals with different coupling.\cite{Roothaan1960,HiraoNakatsuji1973} What we add follows from these ingredients but has not, to our knowledge, been stated in this form. (a) For two UHF determinants, possibly from different particle-number sectors, the numbers of gained and lost spin-orbitals classify the ghost interaction before any calculation. It vanishes at pair rank one. It has a fixed sign at pair rank two and whenever two or more spin-orbitals are only gained or only lost. Otherwise its sign depends on the orbitals, although each pair of spin-orbitals contributes with a fixed sign. For larger ensembles, vanishing pair covariances of the occupations are sufficient for the two constructions to coincide, and necessary in the complete-basis limit. These statements are short corollaries of the quadratic form of the UHF energy; their use is that they can be read from the occupation patterns. (b) The consequences of these pairs for the orbital gradient and for the chemical potential of prescribed SA-UHF ensembles. (c) Calculations on three prescribed ensembles of chemical interest: the hexaamminecobalt(III/II) couple in the occupation patterns of its experimental spin states, a fraction shared between the degenerate frontier orbitals of benzene, and a hole shared between two stacked ethylene molecules. For each, the ghost interaction is reported together with the orbital-relaxation error of the common-orbital reference.

The ghost interaction is the error of FON-UHF relative to a \emph{prescribed} ensemble on common orbitals. It is not an error relative to the exact energy or to the members optimized separately. If the relevant ensemble has statistically independent occupations, FON-UHF is its exact mean-field energy, and $G$ measures instead the error of using a prescribed ensemble. The common-orbital reference itself misses the orbital relaxation of its members, and we report this error alongside $G$ for every system (Sec.~\ref{sec:relax}).

Section~\ref{sec:theory} presents the theory: the two constructions, the ghost interaction and its classification by pair rank, and its consequences for the orbital gradient and the chemical potential. Section~\ref{sec:comp} gives the computational details. Section~\ref{sec:results} reports the three molecular examples and compares the ghost interaction with the relaxation error of the common-orbital reference. Section~\ref{sec:concl} summarizes the conclusions.

\section{Theory}\label{sec:theory}

\subsection{Notation}

Derivations that are not standard are given in the Supporting Information (SI). All one-particle matrices are represented in a fixed real orthonormal basis $\{\chi_\mu\}$, for example the L{\"o}wdin-orthogonalized atomic-orbital basis; a non-orthogonal basis enters only through the overlap matrix. We use real orbitals throughout. Spin-orbitals $\varphi_p$ have spin $\sigma_p\in\{\alpha,\beta\}$ and are orthonormal within each spin. Molecular spin-orbitals are labelled $i,j$ when occupied, $a,b$ when virtual and $p,q,r,s$ when generic; for two determinants $I$ and $J$, occupied and virtual refer to $I$, so that the spin-orbitals lost from $I$ to $J$ are $i,j$ and those gained are $a,b$. The ket $\kets{p}$ is the column of expansion coefficients, $\proj{p}$ the projector onto $\varphi_p$, and $X_{pq}=\langle\varphi_p|\mathbf{X}|\varphi_q\rangle$.

A \emph{spin-resolved matrix} $\mathbf{X}=(\mathbf{X}^\alpha,\mathbf{X}^\beta)$ is an ordered couple of real symmetric matrices, with spin-summed matrix $\mathbf{X}^{\alpha+\beta}=\mathbf{X}^\alpha+\mathbf{X}^\beta$; sums, products and commutators are taken spin by spin, and $\operatorname{rank}\mathbf{X}=\operatorname{rank}\mathbf{X}^\alpha+\operatorname{rank}\mathbf{X}^\beta$. A determinant is represented by its idempotent spin-resolved density matrix $\mathbf{P}_I$, and $\dP_{IJ}\equiv\mathbf{P}_J-\mathbf{P}_I$ ($\dP\equiv\mathbf{P}_2-\mathbf{P}_1$ for two members). A \emph{sector} is the set of states with a given integer electron number. An \emph{ensemble} is a set of determinants $\mathbf{P}_I$ with weights $w_I>0$, $\sum_I w_I=1$, built on one common orbital set; it is \emph{prescribed} when its sectors, spin states and occupation patterns are fixed in advance rather than obtained by minimization. A spin-orbital is \emph{gained} from $I$ to $J$ if it is occupied in $J$ but not in $I$, and \emph{lost} in the converse case; spin-orbitals with the same list of occupations in all members form an \emph{occupation class}. ``Grand-canonical'' is used only in its historical sense, for the fractional-occupation (product) ensemble,\cite{Abdulnur1972,Jorgensen1972,Zerner1989} not for a thermal ensemble. Braces denote sets, for example $\{0,1\}$ for the two values 0 and 1. Square brackets include the end points of an interval and round brackets exclude them: $[0,1]$ contains 0 and 1, whereas $(0,1)$ does not. Energies are in Hartree or milliHartree (mHa).

\subsection{UHF functional and its exact quadratic expansion}

With $(\mu\nu|\lambda\rho)$ the two-electron integrals in chemists' notation and, for symmetric $\mathbf{D}$,
\begin{equation}
[\mathbf{J}(\mathbf{D})]_{\mu\nu}=\sum_{\lambda\rho}(\mu\nu|\lambda\rho)D_{\lambda\rho},\qquad
[\mathbf{K}(\mathbf{D})]_{\mu\nu}=\sum_{\lambda\rho}(\mu\rho|\lambda\nu)D_{\lambda\rho},
\label{eq:JK}
\end{equation}
the UHF functional\cite{PopleNesbet1954,SzaboOstlund} of a spin-resolved density matrix is
\begin{equation}
E[\mathbf{P}]=E_{\mathrm{nuc}}+\operatorname{Tr}[\mathbf{h}\mathbf{P}^{\alpha+\beta}]+\tfrac12\operatorname{Tr}[\mathbf{P}^{\alpha+\beta}\mathbf{J}(\mathbf{P}^{\alpha+\beta})]-\tfrac12\sum_\tau\operatorname{Tr}[\mathbf{P}^\tau\mathbf{K}(\mathbf{P}^\tau)],
\label{eq:EUHF}
\end{equation}
with Fock matrices $\mathbf{F}^\tau=\mathbf{h}+\mathbf{J}(\mathbf{P}^{\alpha+\beta})-\mathbf{K}(\mathbf{P}^\tau)$. Equation~\ref{eq:EUHF} follows from the Slater--Condon rules and defines $E$ on arbitrary symmetric spin-resolved matrices, idempotent or not. With the symmetric bilinear Coulomb-minus-exchange form and the linear map
\begin{equation}
\begin{aligned}
B(\mathbf{X},\mathbf{Y})&=\operatorname{Tr}[\mathbf{X}^{\alpha+\beta}\mathbf{J}(\mathbf{Y}^{\alpha+\beta})]-\sum_\tau\operatorname{Tr}[\mathbf{X}^\tau\mathbf{K}(\mathbf{Y}^\tau)]=\sum_\tau\operatorname{Tr}[\mathbf{X}^\tau\mathbf{L}(\mathbf{Y})^\tau],\\
\mathbf{L}(\mathbf{X})^\tau&=\mathbf{J}(\mathbf{X}^{\alpha+\beta})-\mathbf{K}(\mathbf{X}^\tau),
\end{aligned}
\label{eq:B}
\end{equation}
the two-electron energy is $\tfrac12 B(\mathbf{P},\mathbf{P})$, $\mathbf{F}^\tau[\mathbf{P}+\mathbf{X}]=\mathbf{F}^\tau[\mathbf{P}]+\mathbf{L}(\mathbf{X})^\tau$, and for any symmetric $\mathbf{X}$
\begin{equation}
E[\mathbf{P}+\mathbf{X}]=E[\mathbf{P}]+\sum_\tau\operatorname{Tr}[\mathbf{X}^\tau\mathbf{F}^\tau[\mathbf{P}]]+\tfrac12 B(\mathbf{X},\mathbf{X})
\label{eq:expand}
\end{equation}
with no remainder: the Fock matrix is the first derivative of the energy and $B$ its constant second derivative. Nearly every result below follows from Eq.~\ref{eq:expand}. When $\mathbf{X}$ is a signed sum of projectors, $\mathbf{X}^\tau=\sum_{p:\sigma_p=\tau}s_p\proj{p}$ with $s_p\in\mathbb{R}$,
\begin{equation}
B(\mathbf{X},\mathbf{X})=\sum_{p,q}s_ps_qA_{pq}=2\sum_{p<q}s_ps_qA_{pq},\qquad A_{pq}\equiv J_{pq}-\delta_{\sigma_p\sigma_q}K_{pq},
\label{eq:Apq}
\end{equation}
because $A_{pp}=0$: in HF a spin-orbital does not interact with itself. For $p\neq q$, $A_{pq}>0$. For opposite spins $A_{pq}=J_{pq}$ is a Coulomb repulsion; for equal spins $J_{pq}-K_{pq}$ is the Coulomb repulsion of the two electrons in the antisymmetrized pair state, which is also positive (SI, Sec.~S1.1). The signs of almost all later results follow from this positivity.

\subsection{Ensembles of determinants, SA-UHF and pair rank}

Let $K$ determinants $\mathbf{P}_I^\tau=\sum_r n_r^{\tau,I}\proj{r}^\tau$, $n_r^{\tau,I}\in\{0,1\}$, share one orbital set and carry weights $w_I$. The ensemble density matrix
\begin{equation}
\Pb^\tau=\sum_I w_I\mathbf{P}_I^\tau=\sum_r\bar n_r^\tau\proj{r}^\tau,\qquad \bar n_r^\tau=\sum_I w_In_r^{\tau,I}\in[0,1],
\label{eq:Pbar}
\end{equation}
commutes with every $\mathbf{P}_I$, and $\operatorname{Tr}\Pb^{\alpha+\beta}=\sum_Iw_IN_I=\langle N\rangle$. Rotations within an occupation class leave every $\mathbf{P}_I$ invariant; equal ensemble occupations $\bar n_r=\bar n_s$ do not imply equal occupation patterns. The SA-UHF energy is the weighted average of the determinant energies,
\begin{equation}
\Esa=\sum_I w_IE[\mathbf{P}_I].
\label{eq:ESA}
\end{equation}
When all members belong to one sector, as for the ethylene dimer below, SA-UHF is state-averaged HF with UHF determinants;\cite{Cernatic2022} unlike state-averaged CASSCF, the members are not mixed, and the weights are prescribed rather than ordered as in the GOK principle.\cite{GOK1988}

At fixed orbitals the weights that minimize $\Esa$ at a given $\langle N\rangle$ can be chosen on at most two sectors, adjacent ones when the UHF sector energies satisfy the PPLB convexity condition\cite{Perdew1982} (SI, Sec.~S2.1). Because $\mathbf{J}$ and $\mathbf{K}$ are linear, the ensemble Fock matrix is $\Fb^\tau=\sum_Iw_I\mathbf{F}^\tau[\mathbf{P}_I]=\mathbf{F}^\tau[\Pb]$ for any ensemble.

The fractional complexity of an ensemble is measured by the ensemble rank and by the pair rank of two members,
\begin{equation}
R_{\mathrm{ens}}=\#\{(r,\tau):0<\bar n_r^\tau<1\},\qquad R_{IJ}=\operatorname{rank}(\mathbf{P}_J-\mathbf{P}_I),
\label{eq:rank}
\end{equation}
which coincide for two members with internal weight $0<t<1$. If $m_+$ spin-orbitals $\varphi_a$ are gained and $m_-$ spin-orbitals $\varphi_i$ are lost from $I$ to $J$, then $\mathbf{P}_J-\mathbf{P}_I=\sum_a\proj{a}-\sum_i\proj{i}$ is diagonal with eigenvalues $\pm1$, so that
\begin{equation}
R_{IJ}=m_++m_-,\qquad N_J-N_I=m_+-m_-,\qquad R_{IJ}=|N_J-N_I|+2\min(m_+,m_-).
\label{eq:mpm}
\end{equation}
At fixed electron number $R_{IJ}$ is twice the excitation level. For an adjacent pair ($N_J=N_I+1$) the rank is odd, and it equals one if and only if a single spin-orbital $p$ of spin $\sigma$ is gained, $\mathbf{P}_J^\sigma=\mathbf{P}_I^\sigma+\proj{p}$, $\mathbf{P}_J^{\bar\sigma}=\mathbf{P}_I^{\bar\sigma}$. For $|N_J-N_I|\in\{0,2\}$ and $\mathbf{P}_J\neq\mathbf{P}_I$ the rank is even and at least two, and three or more distinct determinants always contain a pair with $R_{IJ}\geq2$ (SI, Sec.~S2.2). What matters is which spin-orbitals change, not the difference in electron number: a singlet and a quartet on the same orbitals differ by one electron but have $m_+=2$, $m_-=1$ and $R=3$.

\subsection{FON-UHF as an ensemble with independent occupations}

FON-UHF assigns occupations $n_r^\tau\in[0,1]$ to one orbital set and evaluates
\begin{equation}
\Efon=E[\mathbf{P}(\mathbf{n})],\qquad \mathbf{P}^\tau(\mathbf{n})=\sum_r n_r^\tau\proj{r}^\tau.
\label{eq:EFON}
\end{equation}
At fixed occupations the orbitals are stationary when
\begin{equation}
[\mathbf{F}^\tau,\mathbf{P}^\tau]=\mathbf{0},\quad\text{i.e.}\quad \mathbf{FDS}-\mathbf{SDF}=\mathbf{0}\ \text{in the atomic-orbital basis},
\label{eq:FONstat}
\end{equation}
a Roothaan--Hall problem with prescribed occupations. A change $\Delta n$ of one occupation at fixed orbitals changes $\mathbf{P}$ by $\Delta n\proj{r}$, and $B(\mathbf{X},\mathbf{X})=(\Delta n)^2A_{rr}=0$, so Eq.~\ref{eq:expand} gives the HF analogue of Janak's theorem,\cite{Slater1972,Janak1978,Baerends2020} obtained for GC HF in ref~\citenum{Abdulnur1972} and for frozen orbitals in ref~\citenum{Brandi1980},
\begin{equation}
\partial\Efon/\partial n_r^\tau=[\mathbf{F}^\tau]_{rr},
\label{eq:janak}
\end{equation}
with no curvature.

\begin{figure}
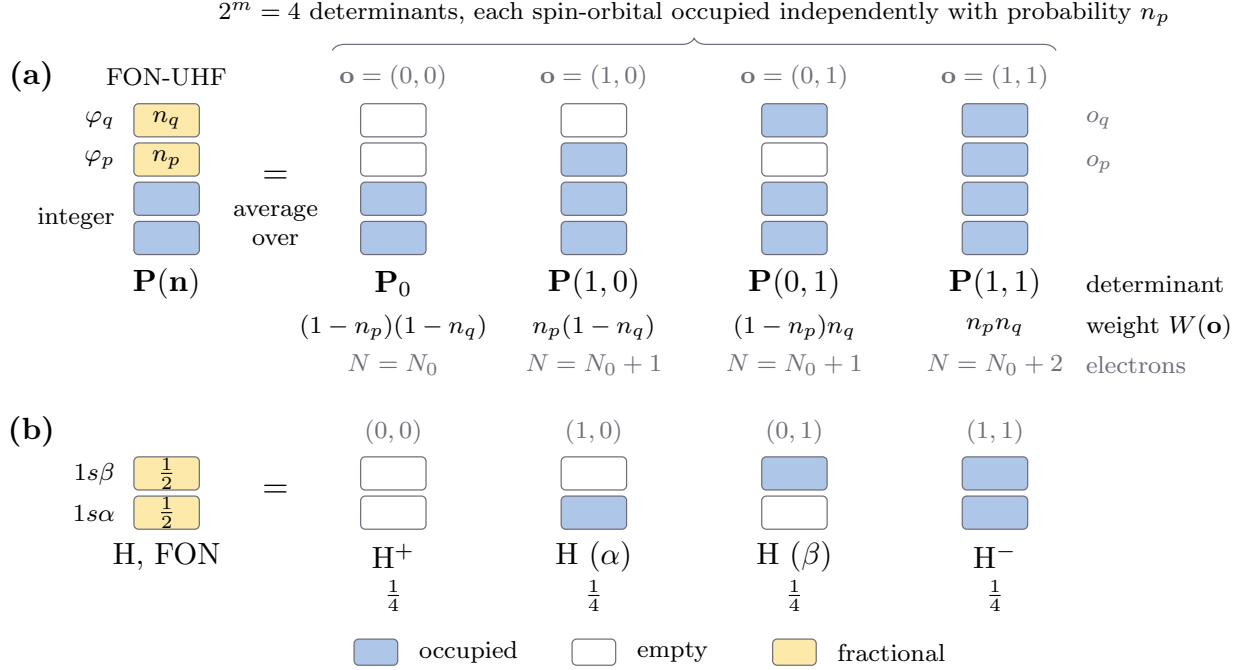

\centering
\resizebox{\linewidth}{!}{\FigIndependent}
\caption{FON-UHF as an ensemble of determinants with independent occupations. (a) Two fractional spin-orbitals $\varphi_p$ and $\varphi_q$ (yellow) on top of fully occupied ones (blue); the four determinants $\mathbf{P}(\mathbf{o})$ carry the weights $W(\mathbf{o})$ of Eq.~\ref{eq:W}, their averaged density matrix is $\mathbf{P}(\mathbf{n})$, and their average energy is $\Efon$, Eq.~\ref{eq:wick}. (b) The fractional-spin hydrogen atom is the ensemble $\tfrac14\mathrm{H}^++\tfrac12\mathrm{H}+\tfrac14\mathrm{H}^-$.}
\label{fig:fon}
\end{figure}

The FON-UHF energy is the average energy of an ensemble of ordinary determinants (Figure~\ref{fig:fon}a). Let $m$ spin-orbitals have fractional occupations $n_p\in(0,1)$, let $\mathbf{P}_0$ keep all integer occupations and leave the $m$ fractional spin-orbitals empty, and let $\mathbf{P}(\mathbf{o})$, $\mathbf{o}\in\{0,1\}^m$, be the $2^m$ determinants obtained by filling them in all possible ways, with weights
\begin{equation}
W(\mathbf{o})=\prod_p n_p^{o_p}(1-n_p)^{1-o_p}.
\label{eq:W}
\end{equation}
Equations~\ref{eq:expand} and \ref{eq:Apq} with $\mathbf{X}=\sum_po_p\proj{p}$ give
\begin{equation}
E[\mathbf{P}(\mathbf{o})]=E[\mathbf{P}_0]+\sum_po_pF^0_{pp}+\sum_{p<q}o_po_qA_{pq},
\label{eq:Eo}
\end{equation}
with $\mathbf{F}^0$ the Fock matrix of $\mathbf{P}_0$ and no $o_p^2$ term because $A_{pp}=0$; the same expression with $n_p$ in place of $o_p$ is $\Efon$. Under $W$, $\langle o_p\rangle=n_p$ and $\langle o_po_q\rangle=n_pn_q$ for $p\neq q$, so averaging Eq.~\ref{eq:Eo} term by term gives
\begin{equation}
\sum_{\mathbf{o}}W(\mathbf{o})\,E[\mathbf{P}(\mathbf{o})]=\Efon.
\label{eq:wick}
\end{equation}
FON-UHF is therefore exactly the average energy of an ensemble with statistically independent occupations, the GC HF ensemble\cite{Abdulnur1972} and the content of Wick's theorem for a quasi-free state with diagonal one-particle density matrix.\cite{Gaudin1960,BLS1994} It spans every sector from $\#\{n_p=1\}$ to $\#\{n_p>0\}$ electrons: the fractional-spin hydrogen atom, $n_{1s\alpha}=n_{1s\beta}=\tfrac12$, is exactly $\tfrac14\mathrm{H}^++\tfrac12\mathrm{H}+\tfrac14\mathrm{H}^-$ (Figure~\ref{fig:fon}b), whose HF fractional-spin error has been analysed in the same terms.\cite{Burton2021}

Moving an amount $s$ of electron from spin-orbital $r$ to $q$ at fixed orbitals gives, by Eqs.~\ref{eq:expand}--\ref{eq:Apq}, the downward-opening parabola
\begin{equation}
\Efon(\mathbf{n}+s(\mathbf{e}_q-\mathbf{e}_r))=\Efon(\mathbf{n})+s(F_{qq}-F_{rr})-s^2A_{qr};
\label{eq:parabola}
\end{equation}
if two occupations are fractional the energy decreases in at least one direction, so a local minimum over occupations has at most one fractional spin-orbital per constrained particle number (refs~\citenum{LiLuYang2017,BLLS1994,Lieb1981}). Several fractional occupations require a prescribed ensemble, or a smearing scheme whose ensemble is that of Eq.~\ref{eq:wick}.

\subsection{The ensemble ghost interaction}

Writing each determinant as the ensemble density matrix plus a finite deviation, $\mathbf{P}_I=\Pb+\Pt_I$ with $\sum_Iw_I\Pt_I=\mathbf{0}$, Eq.~\ref{eq:expand} about $\Pb$ gives $E[\mathbf{P}_I]=E[\Pb]+\sum_\tau\operatorname{Tr}[\Pt_I^\tau\Fb^\tau]+\tfrac12B(\Pt_I,\Pt_I)$, and the linear terms cancel in the weighted sum. Expanding $B$ bilinearly with $B_{IJ}=B(\mathbf{P}_I,\mathbf{P}_J)$,
\begin{equation}
\sum_Iw_IB(\Pt_I,\Pt_I)=\tfrac12\sum_{I,J}w_Iw_JB(\dP_{IJ},\dP_{IJ})=\sum_Iw_IB_{II}-\sum_{I,J}w_Iw_JB_{IJ},
\label{eq:spread}
\end{equation}
so that, for any ensemble, any weights and any orbital set,
\begin{equation}
G\equiv\Efon-\Esa=-\tfrac12\sum_Iw_IB(\Pt_I,\Pt_I)=-\tfrac12\sum_{I<J}w_Iw_J\,B(\dP_{IJ},\dP_{IJ}),
\label{eq:Gpair}
\end{equation}
which for two determinants with weights $(1-t,t)$ reduces to
\begin{equation}
G=-\tfrac12t(1-t)\,B(\dP,\dP).
\label{eq:G2}
\end{equation}
We call $G$ the \emph{ensemble ghost interaction}, with the sign convention of ref~\citenum{Gidopoulos2002PRL}. The middle expression of Eq.~\ref{eq:Gpair} is minus half the spread of the members about their mean, measured with $B$; since $B$ is not positive definite, this spread can have either sign. The last has the form of the pairwise correction of Pastorczak and Pernal,\cite{PastorczakPernal2014} here written for UHF density matrices and for members in different sectors. FON-UHF averages the density matrices before evaluating an energy that is quadratic in them; members that are never present together then interact through the cross terms $w_Iw_JB_{IJ}$, and the interaction within each member is counted with weight $w_I^2$ instead of $w_I$. $\Esa$ contains neither term.

An equivalent form follows from occupations. Equation~\ref{eq:Eo} holds for every member on the common orbitals, with $\mathbf{P}_0$ the determinant in which the fractional spin-orbitals of $\Pb$ are empty. Reading $o_p$ as a random variable that takes the value $n_p^I$ with probability $w_I$, $\langle o_p\rangle=\bar n_p$ and $\langle o_po_q\rangle=\sum_Iw_In_p^In_q^I$ is the probability that $\varphi_p$ and $\varphi_q$ are occupied together. Averaging Eq.~\ref{eq:Eo},
\begin{equation}
\Esa=E[\mathbf{P}_0]+\sum_p\bar n_pF^0_{pp}+\sum_{p<q}\langle o_po_q\rangle A_{pq},
\label{eq:ESAocc}
\end{equation}
whereas $\Efon$ has $\bar n_p\bar n_q$ in place of $\langle o_po_q\rangle$; hence
\begin{equation}
G=-\sum_{p<q}\Cov(o_p,o_q)\,A_{pq},\qquad \Cov(o_p,o_q)=\langle o_po_q\rangle-\bar n_p\bar n_q.
\label{eq:Gcov}
\end{equation}
Only pairs of fractional spin-orbitals contribute. For one electron shared equally between $\varphi_p$ and $\varphi_q$ the members have $(o_p,o_q)=(1,0)$ and $(0,1)$; FON-UHF weights $A_{pq}$ with $\bar n_p\bar n_q=\tfrac14$ although the two are never occupied together, so $\Cov=-\tfrac14$ and $G=\tfrac14A_{pq}$: the electron repels itself across two members that never coexist, exactly the weight of the doubly filled determinant in the independent-occupation ensemble of Eq.~\ref{eq:wick}. Equation~\ref{eq:Gcov} is the general form, for any determinant ensemble across sectors, of the difference between the canonical and grand-canonical functionals of one open subshell,\cite{Abdulnur1972,JorgensenOhrn1973PRA,Zerner1989} of the pair occupations of LEXX\cite{GouldDobson2013} and of the pair-coefficient ghost energy of ref~\citenum{GouldKronik2021} (SI, Sec.~S4.1). $G$ is neither self-interaction, which is the diagonal $A_{pp}=0$ of HF, nor electron correlation, since both energies are averages of single determinants: it is the energy of the assumption that the occupations are uncorrelated in pairs. Since $\Efon$ depends only on $\Pb$, Eq.~\ref{eq:Gcov} also shows that $\Esa$ depends on the pair occupations $\langle o_po_q\rangle$ and is not a function of $\Pb$ alone as soon as two spin-orbitals are fractional, as known from ensemble DFT and density-matrix functional theory.\cite{GouldPittalis2017,GouldKronikPittalis2026,Gilbert1975,Coleman1963,Valone1980}

\begin{figure}
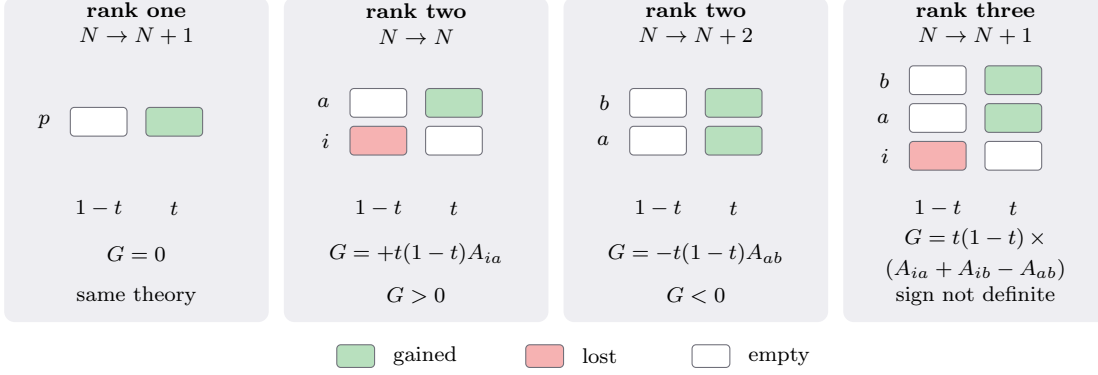

\centering
\FigRank
\caption{Pair rank and ghost interaction for two determinants with weights $1-t$ and $t$; only the spin-orbitals in which the determinants differ are shown. At pair rank one $G=0$ on every orbital set, Eq.~\ref{eq:rank1}. At rank two $G$ is positive for one gained and one lost spin-orbital and negative for two gained or two lost, Eq.~\ref{eq:rank2}. From rank three on, gained--lost pairs contribute with positive sign and gained--gained or lost--lost pairs with negative sign, Eq.~\ref{eq:BdP}; the sum is negative on every orbital set when spin-orbitals are only gained or only lost, and its sign depends on the orbitals otherwise. For spin-orbitals of opposite spin $A_{pq}=J_{pq}$; in the hexaamminecobalt example all gained--lost pairs have opposite spin.}
\label{fig:rank}
\end{figure}

\subsection{Classification by pair rank}\label{sec:class}

For two determinants $\dP=\sum_ps_p\proj{p}$ with $s_p=+1$ on the gained and $-1$ on the lost spin-orbitals, and Eq.~\ref{eq:Apq} gives
\begin{equation}
B(\dP,\dP)=2\Big[\sum_{a<b}A_{ab}+\sum_{i<j}A_{ij}-\sum_{a,i}A_{ai}\Big],
\label{eq:BdP}
\end{equation}
with $a,b$ gained and $i,j$ lost. By Eq.~\ref{eq:G2}, each pair of changing spin-orbitals contributes to $G$ with a fixed sign (Figure~\ref{fig:rank}). A gained--lost pair contributes $+t(1-t)A_{ai}$: FON-UHF lets the two interact although they are never occupied together. A gained--gained or lost--lost pair contributes $-t(1-t)A_{ab}$ or $-t(1-t)A_{ij}$: the two are always occupied together in one member, while FON-UHF counts their repulsion with the product of the occupations. The numbers of gained and lost spin-orbitals then decide the two-member case.

\paragraph{Rank one.} Let $\mathbf{P}_2=\mathbf{P}_1+\proj{p}$ in spin $\sigma$, with weights $(1-t,t)$. Then $\Pb=\mathbf{P}_1+t\dP$ is the FON-UHF density matrix with occupation $t$ of $\varphi_p$, $B(\dP,\dP)=A_{pp}=0$, and Eq.~\ref{eq:expand} gives
\begin{equation}
E[\mathbf{P}_1+t\dP]=E[\mathbf{P}_1]+t[\mathbf{F}_1^\sigma]_{pp}=(1-t)E[\mathbf{P}_1]+tE[\mathbf{P}_2],\qquad\text{i.e.}\quad \Efon=\Esa,\ G=0,
\label{eq:rank1}
\end{equation}
for every $t$ and every common orbital set. The two energies are therefore the same function of the orbitals, and their gradients, Hessians and stationary points coincide; this is the frozen-orbital linearity of HF in one occupation\cite{LiYang2017,Kraisler2013} read as an identity between functionals. The Coulomb and exchange fields of the added electron cancel on its own orbital, $[\mathbf{L}(\proj{p})^\sigma]_{qp}=(qp|pp)-(qp|pp)=0$. The $p$-th column of the spin-$\sigma$ Fock matrix is therefore the same in the two sectors and in $\Fb$, and the chemical potential
\begin{equation}
\mu\equiv E[\mathbf{P}_2]-E[\mathbf{P}_1]=\partial\Efon/\partial n_p^\sigma=[\Fb^\sigma]_{pp}
\label{eq:mu1}
\end{equation}
is independent of $t$ at fixed orbitals and, at stationary orbitals with $0<t<1$, is the eigenvalue $\varepsilon_p$ of $\Fb^\sigma$. Conversely, if $\Pb$ has exactly one fractional occupation, every member has the same occupations except possibly for that spin-orbital, and up to duplicates the ensemble is a rank-one adjacent pair (SI, Sec.~S3.1). With a fraction $c_{\mathrm x}<1$ of exact exchange $A_{pp}=(1-c_{\mathrm x})J_{pp}\neq0$ and the equivalence fails already at rank one; the difference is then fractional self-interaction,\cite{Vydrov2007,MoriSanchez2008PRL,Kraisler2013} not ghost interaction (SI, Sec.~S3.2).

\paragraph{Rank two.} With $\dP=s_p\proj{p}+s_q\proj{q}$, Eqs.~\ref{eq:Apq} and \ref{eq:G2} give
\begin{equation}
G=-s_ps_q\,t(1-t)A_{pq}\neq0
\label{eq:rank2}
\end{equation}
for every orbital set and every $t\in(0,1)$, because $A_{pq}>0$. Pair rank two is therefore never equivalent at common orbitals, with $G>0$ for one gained and one lost spin-orbital (same sector) and $G<0$ for two gained or two lost (a non-adjacent pair, $N_J=N_I\pm2$). This does not by itself order the minima of the two functionals optimized separately, although they differ in every case computed here.

\paragraph{Rank three and higher.} When spin-orbitals are only gained ($m_-=0$) or only lost ($m_+=0$), every term of Eq.~\ref{eq:BdP} enters $G$ with negative sign, and $G=-t(1-t)\sum_{a<b}A_{ab}<0$, or the same sum over lost spin-orbitals, on every orbital set and at any rank. When both are present and $R_{IJ}\geq3$, terms of both signs occur. For $m_+=2$, $m_-=1$, with $a,b$ gained and $i$ lost, Eq.~\ref{eq:BdP} gives $B(\dP,\dP)=2(A_{ab}-A_{ia}-A_{ib})$ and $G=t(1-t)(A_{ia}+A_{ib}-A_{ab})$. This is positive when the three spin-orbitals are compact orbitals on one centre, as in the cobalt pair of Sec.~\ref{sec:cobalt}. It is negative when $a$ and $b$ lie on one fragment and $i$ on a distant one, since $A_{ia}$ and $A_{ib}$ then decay as the inverse distance while $A_{ab}$ stays finite. In between it vanishes for particular orbitals. The same two placements show that, in a complete basis, both signs occur for every $(m_+,m_-)$ with $m_+,m_-\geq1$ and $R_{IJ}\geq3$ (SI, Sec.~S4.3). Beyond rank two, the numbers $(m_+,m_-)$ thus fix the sign of $G$ when one of them is zero, and otherwise only the sign of each pair contribution.

\paragraph{Sign of $G$.} Two limiting cases show that $G$ has no fixed sign. Let a fraction be distributed over several spin-orbitals of the same spin with respect to a common reference, $\mathbf{P}_p=\mathbf{P}_0+\eta\proj{p}$, $\eta=\pm1$, with weights $w_0\geq0$ and $w_p$. Every $\mathbf{P}_p$ forms a rank-one pair with $\mathbf{P}_0$. Two of them differ by one gained and one lost spin-orbital, so $B(\dP_{pq},\dP_{pq})=-2A_{pq}$, and Eq.~\ref{eq:Gpair} gives
\begin{equation}
G=\sum_{p<q}w_pw_qA_{pq}>0
\label{eq:distrib}
\end{equation}
whenever two $w_p$ are positive, consistent with Zerner's observation that GC HF lies above the configuration average.\cite{Zerner1989} For a non-adjacent pair, $\dP=\proj{a}+\proj{b}$ and
\begin{equation}
G=-t(1-t)A_{ab}<0:
\label{eq:nonadj}
\end{equation}
the two electrons are always added together, with weight $t$, but $\Efon$ counts their repulsion with weight $t^2$. Negative contributions of this kind already occur in the LEXX treatment of a fractional $p$ shell, where the total ghost term remains positive.\cite{GouldDobson2013}

\paragraph{Beyond two determinants.} For three or more members at least one pair has rank $\geq2$, yet the contributions of different pairs can cancel. By Eq.~\ref{eq:Gcov}, $G=0$ on every orbital set when $\Cov(o_p,o_q)=0$ for every pair of fractional spin-orbitals. In a complete basis this condition is also necessary, because the integrals $A_{pq}$ of different pairs can then be varied independently of each other. For example, for compact orbitals that do not overlap, $A_{pq}$ is the repulsion of two point charges at their centres, and no combination of these repulsions with non-zero coefficients vanishes for all positions of the centres (SI, Sec.~S4.2). When the covariances do not all vanish, $G$ can therefore vanish only on particular orbital sets. Pairwise, not full, independence suffices: besides the independent-occupation ensemble, an ensemble of four determinants whose six pairs all have rank two satisfies it (SI, Sec.~S4.2). For general ensembles the pair ranks give the sign of each pair's contribution. When every pair of rank two or more contributes with the same fixed sign, as in Eq.~\ref{eq:distrib}, $G$ has that sign on every orbital set; otherwise the pair ranks alone do not decide the sign or the vanishing of $G$, since every ensemble of three or more distinct members contains a pair with $R_{IJ}\geq2$.

\subsection{Orbital gradient and stationarity}\label{sec:grad}

The orbitals are rotated, spin by spin, as
\begin{equation}
\kets{r}(\bm{\kappa})=\sum_s\kets{s}[e^{-\bm\kappa}]_{sr},\qquad \bm\kappa^\top=-\bm\kappa,\qquad e^{-\bm\kappa}\mathbf{P}e^{\bm\kappa}=\mathbf{P}+[\mathbf{P},\bm\kappa]+O(\bm\kappa^2),
\label{eq:rot}
\end{equation}
and Eq.~\ref{eq:expand} gives, for $r<s$,\cite{Helgaker2000}
\begin{equation}
\left.\frac{\partial E[\mathbf{P}(\bm\kappa)]}{\partial\kappa_{sr}}\right|_{\bm\kappa=\mathbf{0}}=2[\mathbf{F},\mathbf{P}]_{rs}.
\label{eq:grad}
\end{equation}
All members rotate together, so the SA-UHF gradient is $2\sum_Iw_I[\mathbf{F}_I,\mathbf{P}_I]$ with $\mathbf{F}_I=\mathbf{F}[\mathbf{P}_I]$. Writing
\begin{equation}
\mathbf{P}_I=\Pb+\Pt_I,\qquad \mathbf{F}_I=\Fb+\mathbf{L}(\Pt_I),\qquad \sum_Iw_I\Pt_I=\mathbf{0},
\label{eq:fluct}
\end{equation}
the cross terms vanish because $\mathbf{L}$ is linear:
\begin{equation}
\mathbf{g}\equiv2\sum_Iw_I[\mathbf{F}_I,\mathbf{P}_I]=\underbrace{2[\Fb,\Pb]}_{\mathbf{g}_{\mathrm{FON}}}+2\sum_Iw_I[\mathbf{L}(\Pt_I),\Pt_I],
\label{eq:gcov}
\end{equation}
which for two determinants becomes
\begin{equation}
\mathbf{g}=\mathbf{g}_{\mathrm{FON}}+2t(1-t)[\mathbf{L}(\dP),\dP].
\label{eq:g2}
\end{equation}
The first term is the gradient of FON-UHF; the second is a covariance between Fock-matrix and density fluctuations. At rank one the $p$-th column of $\mathbf{L}(\dP)$ vanishes, so that
\begin{equation}
\mathbf{L}(\dP)\dP=\dP\mathbf{L}(\dP)=\mathbf{0},
\label{eq:Lzero}
\end{equation}
and the covariance term vanishes identically, as required by Eq.~\ref{eq:rank1}. The condition $\sum_Iw_I[\mathbf{F}_I,\mathbf{P}_I]=\mathbf{0}$ is that of state-averaged common-orbital methods.\cite{Gidopoulos2002PhysB,Cernatic2022,WernerMeyer1981,DockenHinze1972} That it cannot in general be reduced to the diagonalization of one Fock matrix was noted for the canonical functional\cite{Abdulnur1972,JorgensenOhrn1973PRA} and for ensemble and state-averaged HF.\cite{Gidopoulos2002PhysB,Cernatic2022} Beyond rank one, diagonalizing $\Fb$ yields a FON-UHF stationary point, which is in general not an SA-UHF stationary point.

\begin{figure}
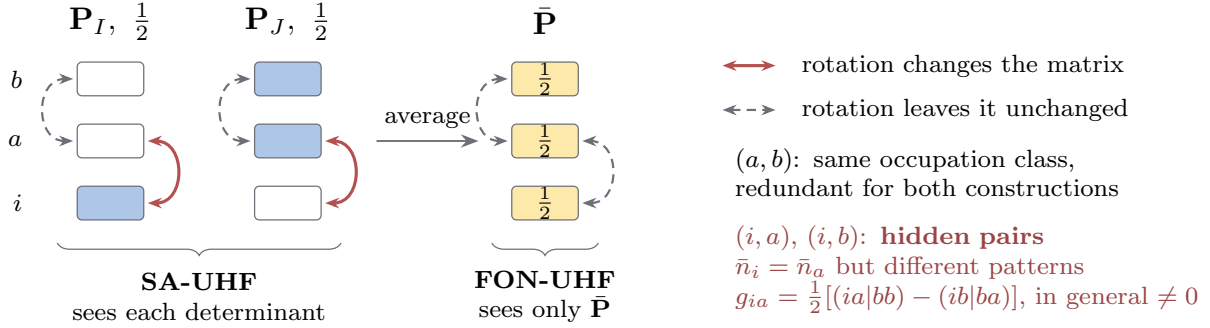

\centering
\resizebox{\linewidth}{!}{\FigHidden}
\caption{Hidden pairs for the rank-three pair $\{i\}\to\{a,b\}$ at $t=\tfrac12$ (equal spins). All three spin-orbitals have $\bar n=\tfrac12$, so no rotation among them changes $\Pb$ and FON-UHF treats all of them as redundant. The rotation $(a,b)$ also leaves both determinants unchanged, whereas $(i,a)$ and $(i,b)$ change $\mathbf{P}_I$ and $\mathbf{P}_J$ and therefore $\Esa$.}
\label{fig:hidden}
\end{figure}

In the common orbital basis every $\mathbf{P}_I$ is diagonal, and the stationarity condition for the pair $(r,s)$ reads
\begin{equation}
g_{rs}=\sum_Iw_I(n_s^I-n_r^I)[\mathbf{F}_I]_{rs}=0,
\label{eq:grs}
\end{equation}
with $g_{rs}$ half the element of $\mathbf{g}$. It holds automatically within an occupation class, where the rotation leaves every $\mathbf{P}_I$ unchanged, and for $\bar n_r\neq\bar n_s$ it is equivalent to
\begin{equation}
F^{\mathrm{eff}}_{rs}\equiv\frac{g_{rs}}{\bar n_s-\bar n_r}=0,
\label{eq:Feff}
\end{equation}
which reduces to $[\Fb]_{rs}=0$ at rank one. For $\bar n_r=\bar n_s$ with different occupation patterns, possible only at $R_{\mathrm{ens}}\geq2$, $g_{rs}=0$ remains a genuine condition while FON-UHF treats the rotation as redundant, because it leaves $\Pb$ unchanged (SI, Sec.~S5.2; Figure~\ref{fig:hidden}). Non-redundant rotations between equally occupied orbitals are familiar from restricted open-shell HF, for open shells with different coupling coefficients.\cite{Roothaan1960,HiraoNakatsuji1973} For such pairs the element
\begin{equation}
\frac{\bar n_r\mathbf{F}_r-\bar n_s\mathbf{F}_s}{\bar n_r-\bar n_s}
\label{eq:coupling}
\end{equation}
of the coupling-operator matrix used in SA-REKS and related ensemble methods is undefined,\cite{Cernatic2022,FilatovShaik1999,Filatov2021} with $\mathbf{F}_r$ the Fock operator weighted by the occupations of $\varphi_r$ in the members. The general coupling operator of Hirao and Nakatsuji remains defined.\cite{HiraoNakatsuji1973} In SA-UHF ensembles such pairs occur between orbitals whose rotation FON-UHF treats as redundant; we call them \emph{hidden pairs}. For a fraction distributed over equivalent orbitals (Eq.~\ref{eq:distrib}) and for the rank-three pair $\{i\}\to\{a,b\}$ at $t=\tfrac12$ (equal spins),
\begin{equation}
g_{pq}=\eta(w_q-w_p)[\mathbf{F}_0]_{pq},\qquad g_{ia}=\tfrac12[(ia|bb)-(ib|ba)]
\label{eq:hidden}
\end{equation}
(SI, Sec.~S5.2). The pair $(p,q)$ is hidden only for $w_p=w_q$, and then $g_{pq}$ vanishes identically: the rotation is an exact zero mode of $\Esa$. The pair $(i,a)$ is a genuine hidden pair, with $g_{ia}$ generally non-zero. A symmetric coupling matrix whose diagonalization has the SA-UHF stationary points as fixed points can always be built (SI, Sec.~S5.3), but it is not unique and its convergence is not guaranteed; direct minimization with the exact gradient, Eq.~\ref{eq:gcov}, is the simplest choice beyond rank one and is used here. In the ensembles of Sec.~\ref{sec:results}, all hidden pairs are exact zero modes of the first type in Eq.~\ref{eq:hidden}. In the rank-three cobalt pair the lost spin-orbital has the other spin, so that no rotation $(i,a)$ exists. The remaining hidden pairs join spin-orbitals that share a fraction with equal weights. For two members, a hidden pair with non-vanishing $g_{rs}$ requires $t=\tfrac12$, a lost and a gained spin-orbital of the same spin and at least one further spin-orbital that changes, that is, a pair rank of at least three.

\subsection{Chemical potential}\label{sec:mu}

Let the weights depend linearly on a parameter $t$, $w_I(t)=w_I^{(0)}+b_It$ with $\sum_Ib_I=0$. At fixed orbitals $\Esa$ is linear in $t$, and its slope defines the chemical potential,
\begin{equation}
\mu=\frac{\mathrm{d}\Esa}{\mathrm{d}t}=\sum_Ib_IE[\mathbf{P}_I].
\label{eq:mu}
\end{equation}
By Eq.~\ref{eq:expand}, the slope of $\Efon(t)=E[\Pb(t)]$ is instead
\begin{equation}
\frac{\mathrm{d}\Efon}{\mathrm{d}t}=\sum_\tau\operatorname{Tr}\Big[\frac{\mathrm{d}\Pb^\tau}{\mathrm{d}t}\,\Fb^\tau\Big],\qquad \frac{\mathrm{d}\Pb}{\mathrm{d}t}=\sum_Ib_I\mathbf{P}_I.
\label{eq:dEFON}
\end{equation}
Differentiating $\Efon=\Esa+G$ gives
\begin{equation}
\mu=\sum_\tau\operatorname{Tr}\Big[\frac{\mathrm{d}\Pb^\tau}{\mathrm{d}t}\,\Fb^\tau\Big]-\left.\frac{\partial G}{\partial t}\right|_{\bm\kappa},\qquad
\left.\frac{\partial G}{\partial t}\right|_{\bm\kappa}=-\tfrac12\sum_{I<J}\big(b_Iw_J+w_Ib_J\big)B(\dP_{IJ},\dP_{IJ}).
\label{eq:mudecomp}
\end{equation}
The first term is the orbital-energy (Janak-type) estimate built from the ensemble Fock matrix, Eq.~\ref{eq:janak} applied to occupations that change together; the second is its error, and it is a direct corollary of Eq.~\ref{eq:Gpair}. Beyond rank one the SA-UHF orbitals do not diagonalize $\Fb$ (Sec.~\ref{sec:grad}), so the diagonal elements that enter the first term are not eigenvalues; for the localized cobalt ensemble of Sec.~\ref{sec:cobalt} at $t=0.25$--0.75 the largest element of $\Fb$ between a fractional and an integer spin-orbital is 0.04--0.11~Hartree. We keep the name orbital-energy estimate because the first term reduces to a signed sum of orbital energies at rank one and on the FON-UHF orbitals. For two determinants with $N_2=N_1+1$ and $\dP=\sum_ps_p\proj{p}$, $\mu=E[\mathbf{P}_2]-E[\mathbf{P}_1]$ and Eq.~\ref{eq:mudecomp} becomes
\begin{equation}
\mu=\sum_ps_p[\Fb^{\sigma_p}]_{pp}+\Big(\tfrac12-t\Big)B(\dP,\dP),
\label{eq:mu2}
\end{equation}
the frozen-orbital statement that a $\Delta$SCF energy involving several spin-orbitals differs from the signed sum of their orbital energies by the two-electron term $\tfrac12B$, here split between the two members. At fixed orbitals
\begin{equation}
\Fb=\mathbf{F}_1+t\mathbf{L}(\dP),\qquad \sum_ps_p[\Fb^{\sigma_p}]_{pp}=\sum_ps_p[\mathbf{F}_1^{\sigma_p}]_{pp}+t\,B(\dP,\dP),
\label{eq:Fbt}
\end{equation}
so the first term of Eq.~\ref{eq:mu2} grows as $tB$ and the second decreases by the same amount: $\mu$ itself does not depend on $t$, while the error of the orbital-energy estimate grows linearly to $|B|/2$ at the end points. For two members the error vanishes at rank one and, at any rank, at $t=\tfrac12$, where Slater's transition-state rule\cite{Slater1972} is exact at fixed orbitals; at rank $\geq3$ it also vanishes on orbital sets with $B(\dP,\dP)=0$. For ensembles with more than two members $\partial G/\partial t$ need not vanish at the symmetric point: for the symmetric cobalt ensemble of Sec.~\ref{sec:cobalt} it equals $-\tfrac19\sum_{k<l}\tfrac12B_{kl}\neq0$ at $t=\tfrac12$. That orbital energies are not simple occupation derivatives when pair averages do not factorize was noted for the canonical functional;\cite{Abdulnur1972} for approximate functionals, the effect of the ensemble generalization on the highest occupied eigenvalue was analysed in ref~\citenum{Kraisler2015JCP}.

At fixed orbitals $\Esa$ is linear in $t$. Its minimum over the orbitals, $E^*(t)=\min_{\bm\kappa}\Esa(\bm\kappa,t)$, is therefore concave in $t$ at any rank. For $t=\theta t_1+(1-\theta)t_2$ and a minimizer $\bm\kappa^*$ at $t$,
\begin{equation}
E^*(t)=\theta\Esa(\bm\kappa^*,t_1)+(1-\theta)\Esa(\bm\kappa^*,t_2)\geq\theta E^*(t_1)+(1-\theta)E^*(t_2).
\label{eq:concave}
\end{equation}
When the end points are the global minima of the two prescribed members, $E^*(t)$ lies above the chord joining them, and in any case $\mu(t)$ is non-increasing; at rank one this is the piecewise concavity of HF.\cite{LiYang2017} By the envelope theorem,\cite{MilgromSegal2002}
\begin{equation}
\frac{\mathrm{d}E^*}{\mathrm{d}t}=\mu,
\label{eq:envelope}
\end{equation}
so that integrating $\mu$ over $t$ gives the change of the optimized SA-UHF energy. The decrease of $\mu$ also holds along every branch of local minima (SI, Sec.~S6.1).

The same concavity holds in a parameter $\lambda$ that distributes a fraction between two symmetry-equivalent orbitals. Symmetry and concavity then give
\begin{equation}
E^*(\lambda)=E^*(1-\lambda)\quad\Longrightarrow\quad E^*(\lambda)\geq E^*(0)=E^*(1),
\label{eq:symdil}
\end{equation}
and the fraction localizes. This is the symmetry dilemma of HF,\cite{LykosPratt1963} not a property of the exact energy; for FON-UHF the same follows from Eq.~\ref{eq:parabola}, but the cost of sharing differs by $G$. If the two orbitals are degenerate but the two localized solutions are not related by a symmetry operation, as for the real components of a degenerate level of benzene (Sec.~\ref{sec:benzene}), $E^*(0)\neq E^*(1)$ in general; concavity then still places the minimum at the lower end point, and the penalty for sharing is measured from it. The two-member relations of this section are checked with exact rational numbers on a three-orbital model in SI, Sec.~S7.

\section{Computational details}\label{sec:comp}

All calculations were performed with a development version of eT 2.6.0,\cite{eT2020} in which FON-UHF and SA-UHF are implemented as separate methods. For an adjacent pair of rank one, FON-UHF is converged with the standard self-consistent-field solver with DIIS acceleration, which by Eq.~\ref{eq:rank1} also yields SA-UHF stationary points; in Sec.~\ref{sec:rank1res} SA-UHF was then minimized directly from the FON-UHF orbitals as a check. A general ensemble is specified as a list of determinants, each with a weight and its occupied $\alpha$ and $\beta$ orbitals. It is optimized by direct minimization over the rotations between spin-orbitals of different occupation pattern. The minimization uses the exact gradient of Eq.~\ref{eq:gcov}, a limited-memory BFGS update with a diagonal preconditioner and an Armijo line search, and stops at $\max|g_{rs}|<10^{-7}$. For the cobalt complex the threshold is $5\times10^{-7}$: with the integral screening thresholds used, the gradient stagnates near $10^{-7}$ while the energy is constant to $10^{-12}$~Hartree. FON-UHF with several fractional spin-orbitals is optimized in the same way as the corresponding independent-occupation ensemble, whose energy equals $\Efon$ on any orbitals (Eq.~\ref{eq:wick}). Relative to FON-UHF, each SA-UHF member needs only a low-rank update of a reference Fock matrix, which for the cobalt ensembles adds 4--11\% to the orbital densities of a Fock build (SI, Sec.~S5.4). For every ensemble the program evaluates $\Esa$, $\Efon$, $G$ and $B(\dP_{IJ},\dP_{IJ})$ for each pair of members on the same orbitals, so Eq.~\ref{eq:Gpair} is checked in every calculation. An independent PySCF\cite{PySCF2020} implementation of both functionals reproduces the eT energies of benzene and the ethylene dimer to $10^{-9}$~Hartree. The same implementation was used to evaluate the elements of $\Fb$ on the optimized eT ensemble orbitals; $\langle\hat S^2\rangle$ and the energies of the individual members were evaluated by eT at fixed ensemble orbitals.

The results use four energy differences. The deviation of an optimized ensemble energy $E^{\mathrm{opt}}$ (SA-UHF or FON-UHF) from the chord of the separately optimized members, each with its own orbitals, is
\begin{equation}
\Delta E=E^{\mathrm{opt}}-\sum_Iw_IE_I^{\mathrm{opt}}.
\label{eq:dEchord}
\end{equation}
The gap between SA-UHF and FON-UHF, each optimized separately with the same ensemble occupations, is
\begin{equation}
\dEgh=E^{\mathrm{opt}}_{\mathrm{FON}}-E^{\mathrm{opt}}_{\mathrm{SA}},
\label{eq:dEgh}
\end{equation}
whereas $G=\Efon-\Esa$ is evaluated on the optimized SA-UHF orbitals. For a fraction shared between two equivalent orbitals or sites, the penalties for sharing are measured from the optimized SA-UHF ensemble with the fraction localized on one of them (the lower one when the two localized solutions differ),
\begin{equation}
\Delta\Esa=E^{\mathrm{opt}}_{\mathrm{SA}}(\text{shared})-E^{\mathrm{opt}}_{\mathrm{SA}}(\text{localized}),\qquad
\Delta\Efon=E^{\mathrm{opt}}_{\mathrm{FON}}(\text{shared})-E^{\mathrm{opt}}_{\mathrm{SA}}(\text{localized}).
\label{eq:penalty}
\end{equation}
A ``spurious'' repulsion is one that is absent from the prescribed ensemble. Energy differences are given in mHa throughout the tables, total energies in Hartree.

With several fractional spin-orbitals both functionals have several stationary points;\cite{SeegerPople1977,ThomHeadGordon2008} for each ensemble the lowest solution found from the starting orbitals listed in SI, Sec.~S8.2 is reported. For the localized cobalt ensemble, FON-UHF was started from the optimized SA-UHF orbitals at the same $t$. The fractions then occupy the same orbitals, and $\dEgh\leq G$ by construction. At $t=0.25$ PySCF finds a FON-UHF solution 0.57~mHa below the eT one reported here, so the reported $\dEgh$ are upper bounds to the gap with the lowest FON-UHF solution; at this $t$ the difference is 0.4\% of $\dEgh$. For the symmetric cobalt ensemble at $t=0.25$ two SA-UHF solutions were found that differ by 0.14~mHa in $\Esa$ and by 0.8~mHa in $G$; the lower one is reported. Internal (UHF$\to$UHF) stability of the integer-sector determinants was tested with PySCF 2.14 (SI, Table~S4). The orbital Hessians of the optimized SA-UHF and FON-UHF ensembles of benzene and the ethylene dimer, and of the localized rank-three cobalt ensemble at $t=\tfrac12$, have only positive eigenvalues, apart from the exact zero mode of Eq.~\ref{eq:hidden} (SI, Sec.~S8.3; for the cobalt ensemble the Hessian was evaluated with density fitting). For the benzene cation the two real $e_{1g}$ components give two distinct localized solutions; the lower one, which is internally stable, is the reference used here (SI, Sec.~S8.3). Basis sets are cc-pVDZ, with cc-pVTZ for the basis-set check,\cite{Dunning1989} and, for the cobalt complex, def2-SVP.\cite{Weigend2005} For benzene and the ethylene dimer, cc-pVTZ changes $G$ by at most 3.4\% and its share of the FON-UHF penalty by less than one percentage point (SI, Sec.~S8.5). For the cobalt complex, def2-TZVPP on Co with def2-SVP on N and H changes $G$ of the localized ensemble at $t=\tfrac12$ by 0.7\%, the pair integrals $A_{pq}$ by at most 1.4\% and $\dEgh$ by 0.7\% (SI, Sec.~S8.5); the cobalt numbers reported are def2-SVP values for the stationary solutions stated.

\section{Results}\label{sec:results}

\subsection{Rank one}\label{sec:rank1res}

At pair rank one the identities of Sec.~\ref{sec:class} were verified on twelve ensembles of \ce{H2} and \ce{O2} between adjacent sectors: the FON-UHF orbitals are already stationary for $\Esa$, the two energies agree to $10^{-12}$~Hartree, and $\mu=[\Fb^\sigma]_{pp}$ agrees with the numerical derivative $\mathrm{d}E/\mathrm{d}N$ (SI, Sec.~S8.1). The results below concern prescribed ensembles beyond rank one.

\begin{figure}
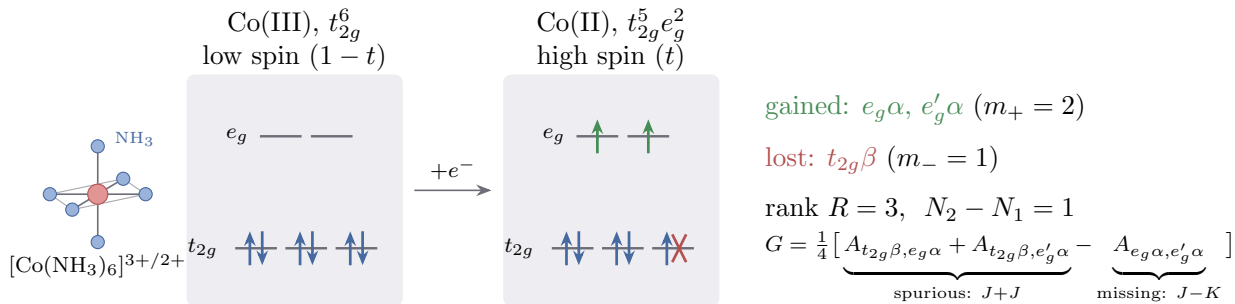

\centering
\resizebox{\linewidth}{!}{\FigCobalt}
\caption{The rank-three redox pair of hexaamminecobalt. With respect to low-spin Co(III), high-spin Co(II) gains the spin-orbitals $e_g\alpha$, $e_g'\alpha$ (green) and loses one $t_{2g}\beta$ (crossed). The two gained--lost pairs have opposite spins ($A=J$), the gained--gained pair equal spins ($A=J-K$).}
\label{fig:co}
\end{figure}

\subsection{A spin-changing redox couple: hexaamminecobalt(III/II)}\label{sec:cobalt}

Hexaamminecobalt(III/II) is a redox couple whose members differ in spin state: low-spin Co(III) ($t_{2g}^6$, $^1A_{1g}$) and high-spin Co(II) ($t_{2g}^5e_g^2$, $^4T_{1g}$). Its self-exchange is slow, about 100 times slower than that of \ce{[Co(en)3]^{3+/2+}}.\cite{Hammershoi1984} Adding one electron gains two $\alpha$ $e_g$ spin-orbitals and loses one $\beta$ $t_{2g}$ (Figure~\ref{fig:co}), so $m_+=2$, $m_-=1$ and the pair rank is three. We use one $O_h$ \ce{CoN6} geometry, intermediate between those of the Co(III) and Co(II) hexaammine salts,\cite{KimeIbers1969} and the def2-SVP basis (SI, Secs.~S8.2 and S9).

UHF favours high spin, as expected for exact exchange:\cite{Reiher2001,Harvey2004} high-spin Co(II) lies 27.6~kcal/mol below low-spin Co(II), but the Co(III) quintet also lies 24.2~kcal/mol below the low-spin singlet (Table~\ref{tab:codet}; SI, Sec.~S8.6), so the UHF ground states would form a rank-one pair. We keep the experimental spin states: the ensemble is prescribed, and its Co(III) member is an excited UHF state. The closed-shell singlet is internally stable; a separate broken-symmetry singlet 2.0~mHa lower would raise every deviation from the chord by $2.0(1-t)$~mHa and leave $G$ unchanged. On the common orbitals the Co(III) member is spin-polarized ($\langle\hat S^2\rangle=0.30$ at $t=\tfrac12$; SI, Sec.~S8.7), so ``low-spin Co(III)'' denotes the occupation pattern $N_\alpha=N_\beta=42$; this enters the deviation from the chord, not $G$. The isolated determinants were optimized from both Cartesian orientations of the degenerate sets and the ensembles from generic orbitals, because ensembles started from a symmetric orientation of the hole stall 1.5--5.8~mHa higher (SI, Sec.~S8.2); in the optimized localized ensemble the hole occupies one spin-orbital, close to $(d_{xy}-d_{yz}-d_{xz})/\sqrt3$ (SI, Sec.~S8.8).

\begin{table}
\caption{\ce{[Co(NH3)6]^{3+/2+}} (def2-SVP): integer-sector determinants in the prescribed spin states, and decomposition of $G$ at $t=\tfrac12$ on the optimized SA-UHF orbitals of the localized rank-three ensemble. $E+1717$ is the total UHF energy shifted by $+1717$~Hartree; $J$, $K$, $A$ and the contributions to $G$ in Hartree.}
\label{tab:codet}
\centering
\begin{tabular}{llrr}
\toprule
Determinant & Configuration & $E+1717$ & $\langle\hat S^2\rangle$\\
\midrule
Co(III), low spin & $t_{2g}^6$ & $-0.188872$ & 0.000\\
Co(II), high spin & $t_{2g}^5e_g^2$ & $-0.790385$ & 3.757\\
Co(II), low spin & $t_{2g}^6e_g^1$ & $-0.746349$ & 1.323\\
\bottomrule
\end{tabular}

\medskip
\begin{tabular}{llrrrr}
\toprule
Pair & Type & $J$ & $K$ & $A$ & Contribution to $G$\\
\midrule
$t_{2g}\beta$--$e_g\alpha$ & gained--lost & 0.7617 & -- & 0.7617 & $+0.1904$\\
$t_{2g}\beta$--$e_g'\alpha$ & gained--lost & 0.7616 & -- & 0.7616 & $+0.1904$\\
$e_g\alpha$--$e_g'\alpha$ & gained--gained & 0.6101 & 0.0289 & 0.5813 & $-0.1453$\\
Total & & & & & 0.2355\\
\bottomrule
\end{tabular}
\end{table}

\begin{table}
\caption{\ce{[Co(NH3)6]^{3+/2+}} (def2-SVP): optimized ensembles $(1-t)\,\mathrm{Co(III)}_{\mathrm{LS}}+t\,\mathrm{Co(II)}_{\mathrm{HS}}$ (rank three, localized hole) and $(1-t)\,\mathrm{Co(III)}_{\mathrm{LS}}+t\,\mathrm{Co(II)}_{\mathrm{LS}}$ (rank one). $\Esa+1717$ in Hartree; $\Delta E$ (deviation from the chord), $G$ and $\dEgh$ in mHa. $\dEgh$ uses FON-UHF started from the SA-UHF orbitals and is an upper bound to the gap with the lowest FON-UHF solution.}
\label{tab:co}
\centering
\begin{tabular}{lrrrrr}
\toprule
 & $t$ & $\Esa+1717$ & $\Delta E$ & $G$ & $\dEgh$\\
\midrule
rank three & 0.25 & $-0.312687$ & 26.56 & 181.1 & 157.2\\
 & 0.50 & $-0.457787$ & 31.84 & 235.5 & 207.4\\
 & 0.75 & $-0.618088$ & 21.92 & 173.9 & 111.3\\
rank one & 0.50 & $-0.429562$ & 38.05 & 0 & 0\\
\bottomrule
\end{tabular}
\end{table}

At half transfer $G=0.236$~Hartree (148~kcal/mol, Table~\ref{tab:co}). It is large because the three spin-orbitals are compact $d$ orbitals on one centre, with Coulomb integrals of 0.6--0.8~Hartree. The gained--lost terms $t(1-t)J$ have the same quadratic form in the occupation as the Hubbard corrections of DFT$+U$, which are added to penalize fractional occupation of localized $d$ shells.\cite{Cococcioni2005} Here the term is instead the error of FON-UHF, with an unscreened $J$. By Eqs.~\ref{eq:G2} and \ref{eq:BdP}, $G(\tfrac12)=\tfrac14(A_{ia}+A_{ib}-A_{ab})$ (Table~\ref{tab:codet}): the two gained--lost pairs ($A=J$) add a repulsion absent from the ensemble, and the gained--gained pair ($A=J-K$) removes one that it contains, cancelling about 40\% of the first. Optimizing FON-UHF separately reduces the gap only in part, and its ensemble, with $N-1$ to $N+2$ electrons, does not describe the Co(III)/Co(II) couple.

\begin{table}
\caption{\ce{[Co(NH3)6]^{3+/2+}}: chemical potential $\mu=\mathrm{d}\Esa/\mathrm{d}t$ of the optimized ensembles and its two parts at fixed orbitals, Eq.~\ref{eq:mudecomp} (Hartree). The first part is the orbital-energy estimate on the SA-UHF orbitals, the second minus the derivative of $G$; their sum is $\mu$. The last column is the same estimate on the optimized FON-UHF orbitals, which equals $\mathrm{d}\Efon/\mathrm{d}t$ along the FON-UHF branch. For the two-member ensembles the first part is $\sum_ps_p[\Fb^{\sigma_p}]_{pp}$ and the second $(\tfrac12-t)B(\dP,\dP)$, with $B(\dP,\dP)=-1.9316$, $-1.8842$, $-1.8550$ at $t=0.25$, 0.5, 0.75.}
\label{tab:comu}
\centering
\begin{tabular}{lrrrrr}
\toprule
Ensemble & $t$ & estimate (SA) & $-\partial G/\partial t|_{\bm\kappa}$ & $\mu$ & estimate (FON)\\
\midrule
rank three, localized & 0.25 & $-0.0579$ & $-0.4829$ & $-0.5408$ & $-0.0212$\\
 & 0.50 & $-0.6139$ & 0 & $-0.6139$ & $-0.6933$\\
 & 0.75 & $-1.1304$ & $+0.4638$ & $-0.6666$ & $-1.0751$\\
rank one & 0.50 & $-0.5769$ & 0 & $-0.5769$ & $-0.5769$\\
symmetric hole, Eq.~\ref{eq:cosym} & 0.25 & $+0.0879$ & $-0.6157$ & $-0.5278$ & $+0.0629$\\
 & 0.50 & $-0.3144$ & $-0.2848$ & $-0.5992$ & $-0.4056$\\
 & 0.75 & $-0.6717$ & $+0.0171$ & $-0.6546$ & $-0.7050$\\
\bottomrule
\end{tabular}
\end{table}

Table~\ref{tab:comu} splits $\mu=E[\mathbf{P}_2]-E[\mathbf{P}_1]$ according to Eq.~\ref{eq:mu2}. On the SA-UHF orbitals the orbital-energy estimate is off by 0.48 and 0.46~Hartree at $t=0.25$ and 0.75 and exact at $t=\tfrac12$; at fixed orbitals the error would reach $|B|/2\approx0.95$~Hartree at the end points. Between $t=0.25$ and 0.75 the estimate changes by $-1.07$~Hartree, mostly through the $t$-dependence of $\Fb$, and $-\partial G/\partial t$ compensates $+0.95$~Hartree of it; the remaining $-0.13$~Hartree is the decrease of $\mu$ required by concavity, Eq.~\ref{eq:concave}. Integrating $\mu$ with Simpson's rule (Eq.~\ref{eq:envelope}) gives $-0.30526$~Hartree, against $\Esa(0.75)-\Esa(0.25)=-0.30540$~Hartree. On the optimized FON-UHF orbitals the estimate differs from $\mu$ by $+0.52$, $-0.08$ and $-0.41$~Hartree (last column), so changing orbitals does not remove the error. For the rank-one couple $\mu$ is the $e_g$ eigenvalue, Eq.~\ref{eq:mu1}; it differs from the rank-three $\mu$ by 37~mHa at $t=\tfrac12$, mainly the 44~mHa high-spin/low-spin gap of Co(II), and is only indicative because low-spin Co(II) is strongly spin-contaminated ($\langle\hat S^2\rangle=1.32$).

\begin{figure}
\centering
\resizebox{0.92\linewidth}{!}{\FigCoSym}
\caption{(a) The localized ensemble of Table~\ref{tab:co}. (b) The symmetric ensemble, Eq.~\ref{eq:cosym}. Besides the three Co(III)--Co(II)[$k$] pairs of rank three, the Co(II) members form three pairs of rank two among themselves (wavy lines), which move the $\beta$ hole from one $t_{2g}$ orbital to another.}
\label{fig:cosym}
\end{figure}

In the localized ensemble the three $t_{2g}$ orbitals are not treated alike, although the $^4T_{1g}$ state of Co(II) is threefold degenerate in $O_h$; the ammine hydrogens lower the symmetry to (nearly) $D_3$, and the high-spin determinants with a $d_{xy}$ and a trigonal hole differ by only 0.025~mHa (SI, Sec.~S8.8). Equal fractional occupation is the usual way to obtain a symmetric density,\cite{Lehtola2020,PerezMartin2008} so we also built
\begin{equation}
(1-t)\,\mathrm{Co(III)}_{\mathrm{LS}}+\frac t3\sum_{k=1}^3\mathrm{Co(II)}_{\mathrm{HS}}[k],
\label{eq:cosym}
\end{equation}
where $\mathrm{Co(II)}_{\mathrm{HS}}[k]$ has the $\beta$ hole in $t_{2g}$ orbital $k$ (Figure~\ref{fig:cosym}). Besides three Co(III)--Co(II)[$k$] pairs of rank three, with weight products $(1-t)t/3$, the Co(II) members form three gained--lost pairs of rank two, with weight products $(t/3)^2$: the canonical--grand-canonical correction of an open subshell,\cite{Abdulnur1972,JorgensenOhrn1973PRA,Zerner1989} here added to the cross-sector pairs. FON-UHF on this density is an ensemble of $2^5=32$ determinants.

The rank-three part of $G$ stays close to $G$ of the localized ensemble, whereas the rank-two part grows as $t^2$ and at $t=0.75$ nearly doubles $G$; at this $t$, $\dEgh$ is 293~mHa for the symmetric ensemble against 111~mHa for the localized one (Table~\ref{tab:cosym}). A density with the symmetry of the complex thus makes the FON-UHF error larger. With four members $\partial G/\partial t=\tfrac19\sum_{k<l}(-\tfrac12B_{kl})\neq0$ at $t=\tfrac12$ (Eq.~\ref{eq:mudecomp}; $B_{kl}=-1.7091$~Hartree, SI, Table~S10), and the orbital-energy estimate is off by 0.28~Hartree also at half transfer (Table~\ref{tab:comu}).

\begin{table}
\caption{\ce{[Co(NH3)6]^{3+/2+}} (def2-SVP): localized (loc) and symmetric (sym), Eq.~\ref{eq:cosym}, $t_{2g}$ hole. All entries in mHa: $\Esa(\mathrm{sym})-\Esa(\mathrm{loc})$; $G$, with the rank-three and rank-two parts of the symmetric ensemble (sum over the three pairs of each type); $\dEgh$.}
\label{tab:cosym}
\centering
\begin{tabular}{rrrrrrrr}
\toprule
 & & & \multicolumn{3}{c}{$G$ (sym)} & \multicolumn{2}{c}{$\dEgh$}\\
\cmidrule(lr){4-6}\cmidrule(lr){7-8}
$t$ & sym$-$loc & $G$ (loc) & rank 3 & rank 2 & total & loc & sym\\
\midrule
0.25 & 2.20 & 181.1 & 176.3 & 18.2 & 194.5 & 157.2 & 163.8\\
0.50 & 5.72 & 235.5 & 225.9 & 71.2 & 297.1 & 207.4 & 277.0\\
0.75 & 9.13 & 173.9 & 162.7 & 156.3 & 319.0 & 111.3 & 292.7\\
\bottomrule
\end{tabular}
\end{table}

The symmetric SA-UHF energy is 2.2--9.1~mHa higher (11.4~mHa at $t=1$), because one set of orbitals cannot relax fully around a hole in three orbitals at once, the symmetry dilemma of Sec.~\ref{sec:mu}. A symmetric density does not require symmetric orbitals. An equal mixture of rotated images of the localized ensemble, each with its own orbitals, has the symmetry of the complex. It also has the same energy, because the ensemble energy is linear in the weights.\cite{Yang2000,GouldKronikPittalis2026} The six images at $t=\tfrac12$ agree to $10^{-4}$~mHa (SI, Sec.~S8.8). The same principle underlies multiplet-sum and ligand-field DFT.\cite{Ziegler1977,Daul1994} We use the localized hole because it is the lower SA-UHF ensemble and contains only rank-three pairs.

HF is not quantitative for spin-state energetics. The decompositions of $G$ and $\mu$ are exact on the stated orbitals, but the magnitudes depend on geometry, basis set and stationary solution and are not an estimate of the redox potential.

\subsection{A fraction shared between the degenerate frontier orbitals of benzene}\label{sec:benzene}

\begin{figure}
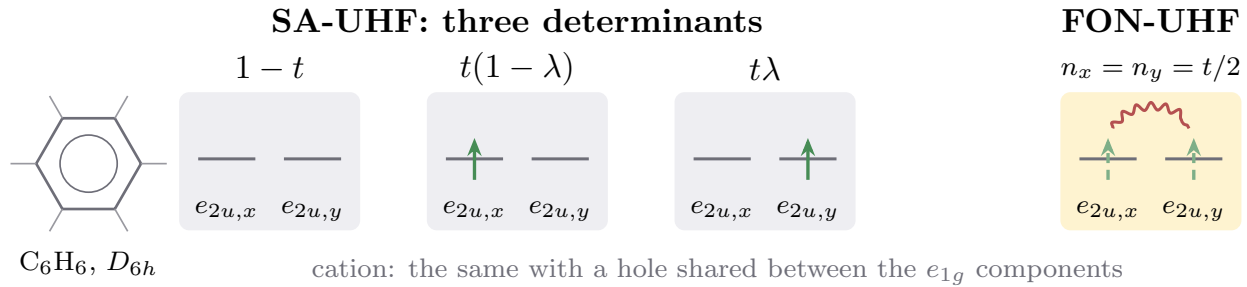

\centering
\resizebox{\linewidth}{!}{\FigBenzene}
\caption{Sharing a fraction of an electron between the degenerate $e_{2u}$ components of benzene. Left: the SA-UHF ensemble, in which the added electron (green) is entirely on one component or entirely on the other. Right: FON-UHF, which gives each component half of the fraction; the two halves repel each other (wavy line).}
\label{fig:bz}
\end{figure}

Both frontier levels of benzene are degenerate: the $e_{1g}$ (HOMO) and $e_{2u}$ (LUMO) $\pi$ orbitals. We share a transferred fraction between the two components of one level (Figure~\ref{fig:bz}). At $N=42.5$ the SA-UHF ensemble is the neutral molecule (weight $1-t$) plus the added electron entirely on one $e_{2u}$ component (weight $t(1-\lambda)$) or on the other (weight $t\lambda$); at $N=41.5$ a hole is shared between the $e_{1g}$ components in the same way. At $\lambda=\tfrac12$ the ensemble density matrix has the $D_{6h}$ symmetry of the nuclear framework, and FON-UHF gives each component the occupation $t/2$ (for the hole, $1-t/2$); at $\lambda=0$ the fraction is localized on one component. The two real components are degenerate, but the UHF solutions with the fraction on one or on the other are not related by a symmetry operation of $D_{6h}$ and have different energies; $\lambda=0$ denotes the lower of the two. We use $t=\tfrac12$, a $D_{6h}$ geometry (SI, Sec.~S9) and cc-pVDZ, without diffuse functions so that the added electron stays in the valence $\pi^*$ orbital.

\begin{table}
\caption{Benzene (cc-pVDZ), $t=\tfrac12$: SA-UHF energy and $G$ as functions of the share $\lambda$, and penalties at $\lambda=\tfrac12$ (mHa); $\lambda=0$ is the lower localized solution. $A_{xy}$ on the optimized FON-UHF and SA-UHF orbitals (Hartree).}
\label{tab:bz}
\centering
\footnotesize
\setlength{\tabcolsep}{3.5pt}
\begin{adjustbox}{max width=\linewidth}\begin{tabular}{lcccccccrr}
\toprule
 & \multicolumn{4}{c}{$\Esa(\lambda)-\Esa(0)$ / $G(\lambda)$} & & & & \multicolumn{2}{c}{$A_{xy}$}\\
\cmidrule(lr){2-5}\cmidrule(lr){9-10}
 & $\lambda=0$ & 0.1 & 0.25 & 0.5 & $\Delta\Efon$ & $\dEgh$ & $G/\Delta\Efon$ & FON & SA\\
\midrule
\ce{C6H6^{-1/2}} ($e_{2u}$) & 0/0 & 1.077/5.536 & 2.231/11.534 & 2.958/15.379 & 18.291 & 15.333 & 84.1\% & 0.2446 & 0.2461\\
\ce{C6H6^{+1/2}} ($e_{1g}$) & 0/0 & 1.652/5.614 & 3.414/11.702 & 4.520/15.607 & 20.115 & 15.595 & 77.6\% & 0.2493 & 0.2497\\
\bottomrule
\end{tabular}\end{adjustbox}
\end{table}

Table~\ref{tab:bz} shows the results. The SA-UHF energy rises monotonically and concavely from $\lambda=0$ to $\lambda=\tfrac12$. The SA-UHF penalty for sharing is 3.0~mHa for the anion and 4.5~mHa for the cation, whereas the FON-UHF penalty is 18--20~mHa, about six times larger for the anion and four and a half times for the cation. Moving the whole fraction to the other component ($\lambda=1$) raises $\Esa$ by 0.20~mHa (anion) and 0.38~mHa (cation). The two members that carry the added electron form a gained--lost pair, and Eq.~\ref{eq:distrib} with $w_x=t(1-\lambda)$ and $w_y=t\lambda$ gives
\begin{equation}
G=t^2\lambda(1-\lambda)A_{xy}>0
\label{eq:Gbz}
\end{equation}
on any common orbital set: in FON-UHF the two halves of the fraction repel each other although they are never present together. On the optimized SA-UHF orbitals $A_{xy}=0.2461$ (anion) and 0.2497~Hartree (cation), 0.6\% and 0.2\% above the values on the FON-UHF orbitals. $G$ and $\dEgh$ each account for 78--84\% of the FON-UHF penalty: FON-UHF attaches about 15~mHa (0.4~eV) to the delocalization of the fraction, almost all of it the repulsion between the two halves. This is the regime where the comparison is cleanest: sharing the fraction on common orbitals costs SA-UHF only 3--4.5~mHa of additional relaxation, against 15--16~mHa of ghost interaction (Table~\ref{tab:bz}).

The isolated anion and cation determinants used for the chord are the lower of the two component solutions and are internally stable. For the cation the other component gives a determinant 2.7~mHa higher that is internally unstable; following its instability leads to the lower one (SI, Table~S4). The isolated anion is a resonance, bound here only because the basis has no diffuse functions, and a Jahn--Teller distortion would lift the degeneracy. The example therefore isolates the effect of the ensemble prescription on a fixed $D_{6h}$ framework and makes no prediction for the relaxed isolated ions.

\subsection{A hole shared between two equivalent centres}\label{sec:dimer}

\begin{figure}
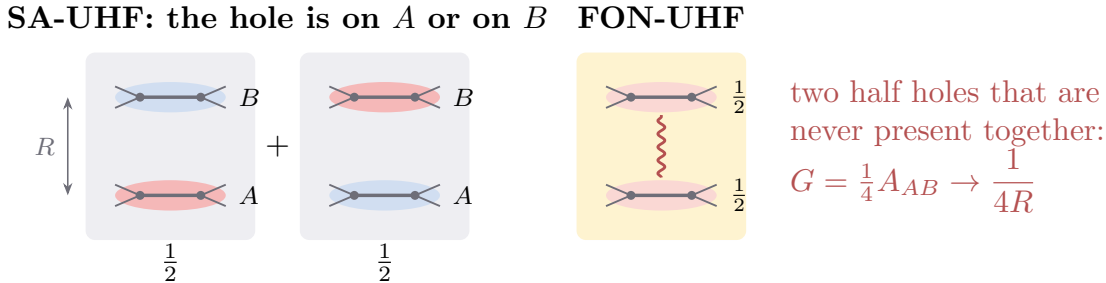

\centering
\resizebox{0.9\linewidth}{!}{\FigDimer}
\caption{A hole (red) shared between two stacked ethylene molecules $A$ and $B$. In SA-UHF the hole is entirely on $A$ or entirely on $B$; in FON-UHF half a hole sits on each monomer, and the two halves repel each other (wavy line).}
\label{fig:dimer}
\end{figure}

The simplest model of a mixed-valence pair\cite{RobinDay1968} is a hole shared between two identical, weakly coupled sites. We place two ethylene molecules face to face at a distance $R$ (geometries in SI, Sec.~S9; cc-pVDZ) and remove one electron. The SA-UHF ensemble $(1-\lambda)\{\text{hole on }\pi_A\}+\lambda\{\text{hole on }\pi_B\}$ is a gained--lost pair, so Eq.~\ref{eq:rank2} gives $G=\lambda(1-\lambda)A_{AB}$, while FON-UHF places half a hole on each $\pi$ orbital (Figure~\ref{fig:dimer}). The members are the integer cation with the hole on one monomer, so the chord is the energy of the localized cation. This is the lowest UHF solution found at all distances considered; at 7 and 10~\AA{} it is internally unstable, but following the instability lowers it by less than 0.01~mHa (SI, Table~S4). At $R=3.5$~\AA{} its Mulliken hole population on one monomer is 1.01, and it lies 3.5~mHa below the symmetric solution. At $R=10$~\AA{} the hole localizes even when the optimization starts from the delocalized HOMO.

\begin{table}
\caption{Ethylene dimer cation (cc-pVDZ), hole shared equally ($\lambda=\tfrac12$): penalties and $G$ (mHa), $A_{AB}$ (Hartree) and $4RG$ ($R$ in bohr).}
\label{tab:dimer}
\centering
\begin{tabular}{rrrrrrr}
\toprule
$R$ (\AA) & $\Delta\Esa$ & $\Delta\Efon$ & $\dEgh$ & $G$ & $A_{AB}$ & $4RG$\\
\midrule
3.5 & 27.593 & 67.732 & 40.138 & 40.726 & 0.1629 & 1.077\\
5.0 & 21.538 & 48.850 & 27.313 & 27.428 & 0.1097 & 1.037\\
7.0 & 21.898 & 41.088 & 19.189 & 19.217 & 0.0769 & 1.017\\
10.0 & 22.098 & 35.423 & 13.325 & 13.331 & 0.0533 & 1.008\\
$\infty$ & 22.216$^a$ & 22.216$^a$ & 0 & 0 & 0 & 1\\
\bottomrule
\end{tabular}

\footnotesize $^a$ $2\Delta E_{\mathrm{chord}}^{\mathrm{mono}}(\tfrac12)$ of a single ethylene, computed separately.
\end{table}

\begin{figure}
\centering
\begin{tikzpicture}
\begin{axis}[width=0.62\linewidth,height=0.42\linewidth,
  xlabel={$1/R$ (bohr$^{-1}$)},ylabel={penalty (mHa)},
  xmin=0,xmax=0.16,ymin=0,ymax=72,
  legend style={at={(1.03,0.5)},anchor=west,font=\footnotesize,draw=none},
  tick label style={font=\footnotesize},label style={font=\small},
  scaled x ticks=false,xtick={0,0.04,0.08,0.12,0.16},
  x tick label style={/pgf/number format/fixed,/pgf/number format/precision=2}]
\addplot[color=red!70!black,mark=square*,mark size=1.6pt] table[x=invR,y=dFON]{dimer.dat};
\addlegendentry{$\Delta E_{\mathrm{FON}}$}
\addplot[color=blue!70!black,mark=o,mark size=1.8pt] table[x=invR,y=dSA]{dimer.dat};
\addlegendentry{$\Delta E_{\mathrm{SA}}$}
\addplot[color=violet,mark=triangle*,mark size=1.8pt] table[x=invR,y=G]{dimer.dat};
\addlegendentry{$G$}
\addplot[domain=0:0.16,dashed,gray,samples=2] {22.216+1000*x/4};
\addlegendentry{$22.216+1/(4R)$}
\addplot[domain=0:0.16,dotted,gray,thick,samples=2] {1000*x/4};
\addlegendentry{$1/(4R)$}
\end{axis}
\end{tikzpicture}
\caption{Ethylene dimer cation: penalties for sharing the hole equally between the two monomers as functions of $1/R$, for $R=3.5$--20~\AA. The ghost interaction approaches the classical repulsion $1/(4R)$ of two half holes, and the FON-UHF penalty approaches $2\Delta E_{\mathrm{chord}}^{\mathrm{mono}}(\tfrac12)+1/(4R)$, Eq.~\ref{eq:asym}; the SA-UHF penalty tends to the constant $2\Delta E_{\mathrm{chord}}^{\mathrm{mono}}(\tfrac12)=22.216$~mHa.}
\label{fig:dimerplot}
\end{figure}
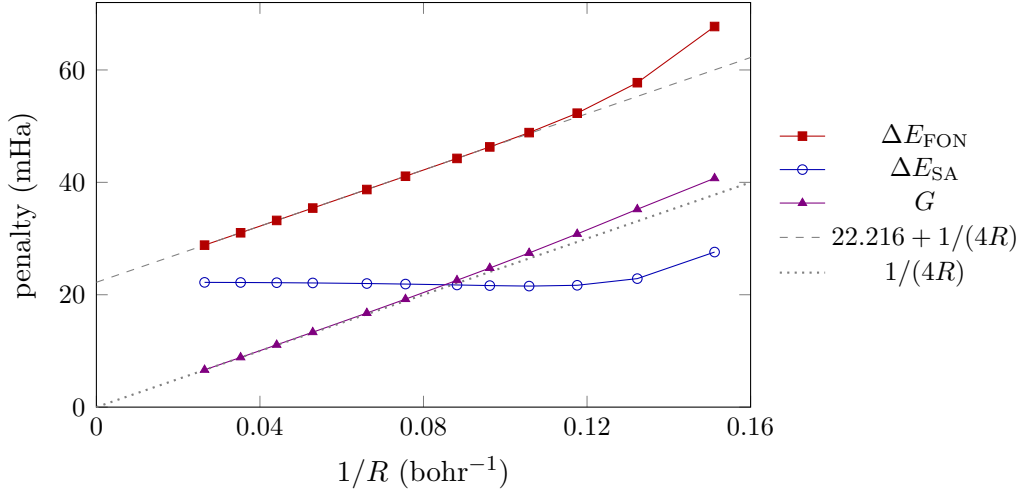

At $\lambda=\tfrac12$, $G=\tfrac14A_{AB}$. As the exchange part dies off, $A_{AB}$ approaches the Coulomb repulsion $1/R$ of two point charges, and $4RG$ is within 8\% of one already at 3.5~\AA{} (Table~\ref{tab:dimer}, Figure~\ref{fig:dimerplot}). The SA-UHF penalty tends instead to a constant. For non-interacting monomers $\Esa(\tfrac12)$ is twice the rank-one ensemble energy of one monomer at $N-\tfrac12$, so $\Delta\Esa$ tends to $2\Delta E_{\mathrm{chord}}^{\mathrm{mono}}(\tfrac12)$, twice the deviation of one monomer from its UHF chord (11.1~mHa for ethylene); at 7 and 10~\AA{} it is within 1.5\% of this limit. The shallow minimum of $\Delta\Esa$ near 5~\AA{} comes from a charge--quadrupole term, $-0.916/R^3$ (atomic units), and a positive induction and overlap remainder, both due to the compromise orbitals of each monomer and unrelated to $G$ (SI, Sec.~S8.4). Thus
\begin{equation}
\Delta\Efon(R)\xrightarrow[R\to\infty]{}2\Delta E_{\mathrm{chord}}^{\mathrm{mono}}(\tfrac12)+\frac1{4R},\qquad
\Delta\Esa(R)\xrightarrow[R\to\infty]{}2\Delta E_{\mathrm{chord}}^{\mathrm{mono}}(\tfrac12).
\label{eq:asym}
\end{equation}

Sharing a hole between infinitely distant sites costs nothing in the exact theory. FON-UHF misses this by the concavity error of the monomers plus the Coulomb tail $1/(4R)$; SA-UHF removes the tail exactly and retains the HF concavity of each fragment, the counterpart of the fractional-charge error of approximate functionals,\cite{MoriSanchez2008PRL,Cohen2008Science} which in Kohn--Sham theory has been removed for dissociated fragments by an ensemble generalization of the functional.\cite{Kraisler2015PRA}

\subsection{Ghost interaction and relaxation error}\label{sec:relax}

The three examples can now be compared with a common measure. The common-orbital ensemble is the reference for $G$ because FON-UHF itself works with one orbital set and one one-particle density matrix, and because only on common orbitals do the two constructions differ by their energy expressions alone. Members optimized separately, each with its own orbitals, give the chord. For the lowest states of two adjacent sectors the chord is the PPLB energy at the UHF level; it is the appropriate reference when no common one-particle density matrix is required. Relative to the chord, the common-orbital ensemble misses orbital relaxation, and FON-UHF adds the ghost interaction on top. Table~\ref{tab:relax} collects the deviations from the chord, Eq.~\ref{eq:dEchord}, and $G$ for every ensemble of this work.

\begin{table}
\caption{Deviations from the chord of the separately optimized members (mHa) of the optimized SA-UHF and FON-UHF energies, and ghost interaction $G$ on the optimized SA-UHF orbitals (mHa), at half transfer ($t=\tfrac12$) or equal sharing ($\lambda=\tfrac12$).}
\label{tab:relax}
\centering
\begin{tabular}{llrrr}
\toprule
System & Ensemble & SA-UHF & FON-UHF & $G$\\
\midrule
\ce{[Co(NH3)6]^{3+/2+}} & rank one, Co(III)$_{\mathrm{LS}}$/Co(II)$_{\mathrm{LS}}$ & 38.05 & 38.05 & 0\\
 & rank three, localized hole & 31.84 & 239.27 & 235.52\\
 & rank three and two, symmetric hole & 37.56 & 314.59 & 297.11\\
\ce{C6H6^{-1/2}} & $\lambda=0$ (rank one) & 9.00 & 9.00 & 0\\
 & $\lambda=\tfrac12$ & 11.96 & 27.29 & 15.38\\
\ce{C6H6^{+1/2}} & $\lambda=0$ (rank one) & 10.82 & 10.82 & 0\\
 & $\lambda=\tfrac12$ & 15.34 & 30.94 & 15.61\\
\ce{(C2H4)2^+} & $R=3.5$~\AA & 27.59 & 67.73 & 40.73\\
 & $R=10$~\AA & 22.10 & 35.42 & 13.33\\
 & $R\to\infty$ & 22.22 & 22.22 & 0\\
\bottomrule
\end{tabular}
\end{table}

In Table~\ref{tab:relax}, the relaxation error of the common-orbital reference, the SA-UHF column, is of the same order in all systems, 9--38~mHa. The ghost interaction ranges from comparable to it (benzene, the dimer at long range) to seven or eight times larger (cobalt). The two errors also differ in kind. For two infinitely separated fragments $\Esa$ is the sum of the fractional-charge ensembles of the fragments, and the relaxation error tends to a constant, twice the concavity error of one fragment (Eq.~\ref{eq:asym}): the common-orbital reference does not reach the chord even at infinite separation. $G$ is instead a long-range Coulomb term that vanishes in this limit.

\section{Conclusions}\label{sec:concl}

For UHF determinants that may belong to different particle-number sectors, FON-UHF and SA-UHF on common orbitals differ by the ensemble ghost interaction. Its pairwise and covariance forms follow in a few lines from the quadratic form of the UHF energy and generalize known results. What the numbers of gained and lost spin-orbitals add is a classification that can be read from the occupations before any calculation. For two determinants, $G$ vanishes on every orbital set at rank one, so that energies, gradients, Hessians, stationary points and chemical potentials coincide. At rank two, $G$ vanishes on no orbital set. Beyond rank two, each gained--lost pair of spin-orbitals adds a repulsion absent from the ensemble, and each gained--gained or lost--lost pair removes one that it contains. The total is then negative on every orbital set when two or more spin-orbitals are only gained or only lost; otherwise its sign depends on the orbitals. For larger ensembles, the two constructions coincide on every orbital set when all pair covariances of the occupations vanish. This condition is also necessary in the complete-basis limit, and the pair ranks alone do not decide it. The same covariance enters the orbital gradient: beyond rank one, diagonalizing the ensemble Fock matrix in general does not give an SA-UHF stationary point, and hidden pairs can appear. At fixed orbitals, its derivative is the error of the orbital-energy estimate of the chemical potential built from the ensemble Fock matrix. For two-member ensembles this error vanishes at rank one, at half transfer and on orbital sets with $B(\dP,\dP)=0$; for larger ensembles it need not vanish at half transfer.

These statements are exact for the UHF functional on any common orbital set. The numbers depend on the configurations, geometries, basis sets and stationary solutions of the three examples. For the hexaamminecobalt(III/II) couple in the occupation patterns of its experimental spin states, a pair of rank three, $G$ is about 0.24~Hartree at half transfer (def2-SVP). The orbital-energy estimate of the chemical potential is off by up to 0.48~Hartree at $t=0.25$ and 0.75 on the SA-UHF orbitals, and by 0.28~Hartree at half transfer for the symmetric four-member ensemble. For a fraction shared between the degenerate frontier orbitals of benzene, $G$ accounts for 78--84\% of the FON-UHF penalty. For a hole shared between two stacked ethylene molecules, $G$ tends to the classical repulsion of two half charges. In all three systems the common-orbital reference carries its own relaxation error of 9--38~mHa, which does not vanish for infinitely separated fragments; $G$ and this error are different quantities and should be reported separately.

Which construction applies is decided by the ensemble. When the occupations of different spin-orbitals are uncorrelated, FON-UHF gives exactly the average UHF energy of that ensemble. For interacting electrons this need not hold: the charging energy would tend to anti-correlate the occupations of near-degenerate orbitals, such as the two frontier components of benzene. The relevant states can also be prescribed, as for the PPLB mixture of two adjacent ground states, a spin-changing redox couple, or a fraction confined by symmetry to one of several equivalent orbitals or sites. The occupations are then in general correlated, and FON-UHF adds the ghost interaction unless the members form a pair of rank one.

The same analysis holds for the Hartree and exact-exchange parts of hybrid functionals, with $A_{pq}$ replaced by $J_{pq}-c_{\mathrm x}\delta_{\sigma_p\sigma_q}K_{pq}$. The diagonal term $(1-c_{\mathrm x})J_{pp}$ then no longer vanishes, so the two constructions differ already at rank one by fractional self-interaction. The off-diagonal terms remain the ghost interaction classified by the pair rank. For correlated wave functions the energy of each state is linear in its two-body reduced density matrix, so averaging the energies of an ensemble and evaluating the energy of the averaged two-body density matrix give the same result. A difference of the kind analysed here appears when a correlated energy is defined directly from fractional orbital occupations, as in fractional-occupation or finite-temperature perturbation theory and in the ensemble-averaged Green's-function extensions of many-body methods to fractional charge.

\section*{Data Availability}
The eT output files, the optimized orbitals, the geometries and the Python scripts, including an independent PySCF implementation used for cross-checks, will be deposited on Zenodo upon publication. The eT implementation will be included in a future public release of eT.

\section*{Author Contributions}
G.F. derived the equations, implemented the methods in eT, performed the calculations, and wrote the manuscript. I.-M.H. acquired the funding and supervised the work.

\begin{acknowledgement}
G.F. and I.-M.H. acknowledge funding from the European Research Council (ERC) under the European Union's Horizon Europe research and innovation programme (ERC Consolidator Grant, grant agreement No.~101170817).
\end{acknowledgement}

\begin{suppinfo}
Derivations of the results on signed projector sums, sector weights, pair rank, the rank-one equivalence, fractional exact exchange, the forms of the ghost interaction and vanishing covariances, the orbital gradient, hidden pairs, the coupling matrix and the chemical potential; a three-orbital model with exact checks; rank-one ensembles of \ce{H2} and \ce{O2}; implementation, occupation patterns, starting orbitals, stability, orbital Hessians, basis-set checks and additional data for the cobalt complex and the ethylene dimer; Cartesian geometries (PDF).
\end{suppinfo}

\bibliography{references}

@article{Abdulnur1972,
  author = {Abdulnur, Suheil F. and Linderberg, Jan and {\"O}hrn, Yngve and Thulstrup, Poul W.},
  title = {Atomic Central-Field Models for Open Shells with Application to Transition Metals},
  journal = {Physical Review A}, year = {1972}, volume = {6}, pages = {889--898},
  doi = {10.1103/PhysRevA.6.889}}

@article{Alam2016,
  author = {Alam, Md. Mehboob and Knecht, Stefan and Fromager, Emmanuel},
  title = {Ghost-Interaction Correction in Ensemble Density-Functional Theory for Excited States with and without Range Separation},
  journal = {Physical Review A}, year = {2016}, volume = {94}, pages = {012511},
  doi = {10.1103/PhysRevA.94.012511}}

@article{BLLS1994,
  author = {Bach, Volker and Lieb, Elliott H. and Loss, Michael and Solovej, Jan Philip},
  title = {There Are No Unfilled Shells in Unrestricted {Hartree--Fock} Theory},
  journal = {Physical Review Letters}, year = {1994}, volume = {72}, pages = {2981--2983},
  doi = {10.1103/PhysRevLett.72.2981}}

@article{BLS1994,
  author = {Bach, Volker and Lieb, Elliott H. and Solovej, Jan Philip},
  title = {Generalized {Hartree--Fock} Theory and the {Hubbard} Model},
  journal = {Journal of Statistical Physics}, year = {1994}, volume = {76}, pages = {3--89},
  doi = {10.1007/BF02188656}}

@article{Baerends2020,
  author = {Baerends, E. J.},
  title = {On derivatives of the energy with respect to total electron number and orbital occupation numbers. {A} critique of {Janak}'s theorem},
  journal = {Molecular Physics}, year = {2020}, volume = {118}, pages = {e1612955},
  doi = {10.1080/00268976.2019.1612955}}

@article{Baiardi2017,
  author = {Baiardi, A. and Paoloni, L. and Barone, V. and Zakrzewski, V. G. and Ortiz, J. V.},
  title = {Assessment of Electron Propagator Methods for the Simulation of Vibrationally Resolved Valence and Core Photoionization Spectra},
  journal = {Journal of Chemical Theory and Computation}, year = {2017}, volume = {13}, pages = {3120--3135},
  doi = {10.1021/acs.jctc.6b00958}}

@article{Brandi1980,
  author = {Brandi, H. S. and De Matos, M. M. and Ferreira, R.},
  title = {Fractional occupation in the {Hartree-Fock} method},
  journal = {Chemical Physics Letters}, year = {1980}, volume = {73}, pages = {597--601},
  doi = {10.1016/0009-2614(80)80726-3}}

@article{Burton2021,
  author = {Burton, Hugh G. A. and Marut, Clotilde and Daas, Timothy J. and Gori-Giorgi, Paola and Loos, Pierre-Fran{\c{c}}ois},
  title = {Variations of the {Hartree--Fock} fractional-spin error for one electron},
  journal = {Journal of Chemical Physics}, year = {2021}, volume = {155}, pages = {054107},
  doi = {10.1063/5.0056968}}

@article{Cernatic2022,
  author = {Cernatic, Filip and Senjean, Bruno and Robert, Vincent and Fromager, Emmanuel},
  title = {Ensemble Density Functional Theory of Neutral and Charged Excitations: Exact Formulations, Standard Approximations, and Open Questions},
  journal = {Topics in Current Chemistry}, year = {2022}, volume = {380}, pages = {4},
  doi = {10.1007/s41061-021-00359-1}}

@article{Chai2012,
  author = {Chai, Jeng-Da},
  title = {Density functional theory with fractional orbital occupations},
  journal = {The Journal of Chemical Physics}, year = {2012}, volume = {136}, pages = {154104},
  doi = {10.1063/1.3703894}}

@article{Cococcioni2005,
  author = {Cococcioni, Matteo and de Gironcoli, Stefano},
  title = {Linear response approach to the calculation of the effective interaction parameters in the {LDA+U} method},
  journal = {Physical Review B}, year = {2005}, volume = {71}, pages = {035105},
  doi = {10.1103/PhysRevB.71.035105}}

@article{Cohen2008FracSpin,
  author = {Cohen, Aron J. and Mori-S{\'a}nchez, Paula and Yang, Weitao},
  title = {Fractional spins and static correlation error in density functional theory},
  journal = {Journal of Chemical Physics}, year = {2008}, volume = {129}, pages = {121104},
  doi = {10.1063/1.2987202}}

@article{Cohen2008Science,
  author = {Cohen, Aron J. and Mori-S{\'a}nchez, Paula and Yang, Weitao},
  title = {Insights into Current Limitations of Density Functional Theory},
  journal = {Science}, year = {2008}, volume = {321}, pages = {792--794},
  doi = {10.1126/science.1158722}}

@article{Coleman1963,
  author = {Coleman, A. J.},
  title = {Structure of Fermion Density Matrices},
  journal = {Reviews of Modern Physics}, year = {1963}, volume = {35}, pages = {668--686},
  doi = {10.1103/RevModPhys.35.668}}

@article{Daul1994,
  author = {Daul, Claude},
  title = {Density functional theory applied to the excited states of coordination compounds},
  journal = {International Journal of Quantum Chemistry}, year = {1994}, volume = {52}, pages = {867--877},
  doi = {10.1002/qua.560520414}}

@article{DockenHinze1972,
  author = {Docken, Kate K. and Hinze, Juergen},
  title = {{LiH} Potential Curves and Wavefunctions for {X$^1\Sigma^+$, A$^1\Sigma^+$, B$^1\Pi$, $^3\Sigma^+$}, and {$^3\Pi$}},
  journal = {Journal of Chemical Physics}, year = {1972}, volume = {57}, pages = {4928--4936},
  doi = {10.1063/1.1678164}}

@article{Dunning1989,
  author = {Dunning, Jr., Thom H.},
  title = {Gaussian basis sets for use in correlated molecular calculations. {I}. The atoms boron through neon and hydrogen},
  journal = {Journal of Chemical Physics}, year = {1989}, volume = {90}, pages = {1007--1023},
  doi = {10.1063/1.456153}}

@article{Filatov2021,
  author = {Filatov, Michael and Lee, Seunghoon and Choi, Cheol Ho},
  title = {Description of Sudden Polarization in the Excited Electronic States with an Ensemble Density Functional Theory Method},
  journal = {Journal of Chemical Theory and Computation}, year = {2021}, volume = {17}, pages = {5123--5139},
  doi = {10.1021/acs.jctc.1c00479}}

@article{FilatovShaik1999,
  author = {Filatov, Michael and Shaik, Sason},
  title = {A spin-restricted ensemble-referenced {Kohn--Sham} method and its application to diradicaloid situations},
  journal = {Chemical Physics Letters}, year = {1999}, volume = {304}, pages = {429--437},
  doi = {10.1016/S0009-2614(99)00336-X}}

@article{GOK1988,
  author = {Gross, E. K. U. and Oliveira, L. N. and Kohn, W.},
  title = {{Rayleigh--Ritz} variational principle for ensembles of fractionally occupied states},
  journal = {Physical Review A}, year = {1988}, volume = {37}, pages = {2805--2808},
  doi = {10.1103/PhysRevA.37.2805}}

@article{Gaudin1960,
  author = {Gaudin, M.},
  title = {Une d{\'e}monstration simplifi{\'e}e du th{\'e}or{\`e}me de {Wick} en m{\'e}canique statistique},
  journal = {Nuclear Physics}, year = {1960}, volume = {15}, pages = {89--91},
  doi = {10.1016/0029-5582(60)90285-6}}

@article{Gidopoulos2002PRL,
  author = {Gidopoulos, N. I. and Papaconstantinou, P. G. and Gross, E. K. U.},
  title = {Spurious Interactions, and Their Correction, in the Ensemble-{Kohn-Sham} Scheme for Excited States},
  journal = {Physical Review Letters}, year = {2002}, volume = {88}, pages = {033003},
  doi = {10.1103/PhysRevLett.88.033003}}

@article{Gidopoulos2002PhysB,
  author = {Gidopoulos, N. I. and Papaconstantinou, P. G. and Gross, E. K. U.},
  title = {Ensemble-{Hartree--Fock} scheme for excited states. The optimized effective potential method},
  journal = {Physica B}, year = {2002}, volume = {318}, pages = {328--332},
  doi = {10.1016/S0921-4526(02)00799-8}}

@article{Gilbert1975,
  author = {Gilbert, T. L.},
  title = {Hohenberg-{Kohn} theorem for nonlocal external potentials},
  journal = {Physical Review B}, year = {1975}, volume = {12}, pages = {2111--2120},
  doi = {10.1103/PhysRevB.12.2111}}

@article{GouldDobson2013,
  author = {Gould, Tim and Dobson, John F.},
  title = {The flexible nature of exchange, correlation, and {Hartree} physics: Resolving ``delocalization'' errors in a ``correlation free'' density functional},
  journal = {Journal of Chemical Physics}, year = {2013}, volume = {138}, pages = {014103},
  doi = {10.1063/1.4773284}}

@article{GouldKronik2021,
  author = {Gould, Tim and Kronik, Leeor},
  title = {Ensemble generalized {Kohn--Sham} theory: The good, the bad, and the ugly},
  journal = {Journal of Chemical Physics}, year = {2021}, volume = {154}, pages = {094125},
  doi = {10.1063/5.0040447}}

@article{GouldKronikPittalis2026,
  author = {Gould, Tim and Kronik, Leeor and Pittalis, Stefano},
  title = {``Ensemblization'' of density functional theory},
  journal = {Journal of Chemical Physics}, year = {2026}, volume = {164}, pages = {040901},
  doi = {10.1063/5.0274509}}

@article{GouldPittalis2017,
  author = {Gould, Tim and Pittalis, Stefano},
  title = {Hartree and Exchange in Ensemble Density Functional Theory: Avoiding the Nonuniqueness Disaster},
  journal = {Physical Review Letters}, year = {2017}, volume = {119}, pages = {243001},
  doi = {10.1103/PhysRevLett.119.243001}}

@article{GrimmeHansen2015,
  author = {Grimme, Stefan and Hansen, Andreas},
  title = {A Practicable Real-Space Measure and Visualization of Static Electron-Correlation Effects},
  journal = {Angewandte Chemie International Edition}, year = {2015}, volume = {54}, pages = {12308--12313},
  doi = {10.1002/anie.201501887}}

@article{GuHirata2024,
  author = {Gu, Pinhao and Hirata, So},
  title = {Thermal mean-field theories},
  journal = {Journal of Chemical Physics}, year = {2024}, volume = {161}, pages = {214108},
  doi = {10.1063/5.0235259}}

@article{Hammershoi1984,
  author = {Hammersh{\o}i, Anders and Geselowitz, Daniel and Taube, Henry},
  title = {Redetermination of the hexaamminecobalt({III}/{II}) electron-self-exchange rate},
  journal = {Inorganic Chemistry}, year = {1984}, volume = {23}, pages = {979--982},
  doi = {10.1021/ic00175a036}}

@incollection{Harvey2004,
  author = {Harvey, J. N.},
  title = {{DFT} Computation of Relative Spin-State Energetics of Transition Metal Compounds},
  booktitle = {Principles and Applications of Density Functional Theory in Inorganic Chemistry {I}},
  series = {Structure and Bonding},
  publisher = {Springer}, address = {Berlin}, year = {2004}, volume = {112}, pages = {151--183},
  doi = {10.1007/b97939}}

@book{Helgaker2000,
  author = {Helgaker, Trygve and J{\o}rgensen, Poul and Olsen, Jeppe},
  title = {Molecular Electronic-Structure Theory},
  publisher = {Wiley}, address = {Chichester}, year = {2000}}

@article{HiraoNakatsuji1973,
  author = {Hirao, Kimihiko and Nakatsuji, Hiroshi},
  title = {General {SCF} operator satisfying correct variational condition},
  journal = {Journal of Chemical Physics}, year = {1973}, volume = {59}, pages = {1457--1462},
  doi = {10.1063/1.1680203}}

@article{JahnTeller1937,
  author = {Jahn, H. A. and Teller, E.},
  title = {Stability of Polyatomic Molecules in Degenerate Electronic States. {I}. Orbital Degeneracy},
  journal = {Proceedings of the Royal Society of London A}, year = {1937}, volume = {161}, pages = {220--235},
  doi = {10.1098/rspa.1937.0142}}

@article{Janak1978,
  author = {Janak, J. F.},
  title = {Proof that $\partial {E}/\partial n_i=\varepsilon_i$ in density-functional theory},
  journal = {Physical Review B}, year = {1978}, volume = {18}, pages = {7165--7168},
  doi = {10.1103/PhysRevB.18.7165}}

@article{Jorgensen1972,
  author = {J{\o}rgensen, Poul},
  title = {Electronic Excitations of Open Shell Systems in the Grand Canonical and Canonical Time-Dependent {Hartree-Fock} Models. Applications on Hydrocarbon Radical Ions},
  journal = {Journal of Chemical Physics}, year = {1972}, volume = {57}, pages = {4884--4892},
  doi = {10.1063/1.1678159}}

@article{JorgensenOhrn1973CPL,
  author = {J{\o}rgensen, Poul and {\"O}hrn, Yngve},
  title = {Triplet--triplet absorption spectra of alternant hydrocarbons within the grand canonical time-dependent {Hartree-Fock} approximation},
  journal = {Chemical Physics Letters}, year = {1973}, volume = {18}, pages = {261--267},
  doi = {10.1016/0009-2614(73)80433-6}}

@article{JorgensenOhrn1973PRA,
  author = {J{\o}rgensen, Poul and {\"O}hrn, Yngve},
  title = {Comparison of Two Statistical Approaches to Calculate Atomic and Molecular Orbitals},
  journal = {Physical Review A}, year = {1973}, volume = {8}, pages = {112--119},
  doi = {10.1103/PhysRevA.8.112}}

@article{KimeIbers1969,
  author = {Kime, N. E. and Ibers, James A.},
  title = {The {Co(III)--N} bond length in relation to the {Co(II)--N} bond length. The crystal structure of hexaamminecobalt({III}) iodide, [{Co(NH$_3$)$_6$}]{I}$_3$},
  journal = {Acta Crystallographica Section B}, year = {1969}, volume = {25}, pages = {168--169},
  doi = {10.1107/S0567740869001968}}

@article{Kraisler2013,
  author = {Kraisler, Eli and Kronik, Leeor},
  title = {Piecewise Linearity of Approximate Density Functionals Revisited: Implications for Frontier Orbital Energies},
  journal = {Physical Review Letters}, year = {2013}, volume = {110}, pages = {126403},
  doi = {10.1103/PhysRevLett.110.126403}}

@article{Kraisler2014,
  author = {Kraisler, Eli and Kronik, Leeor},
  title = {Fundamental gaps with approximate density functionals: The derivative discontinuity revealed from ensemble considerations},
  journal = {Journal of Chemical Physics}, year = {2014}, volume = {140}, pages = {18A540},
  doi = {10.1063/1.4871462}}

@article{Kraisler2015JCP,
  author = {Kraisler, Eli and Schmidt, Tobias and K{\"u}mmel, Stephan and Kronik, Leeor},
  title = {Effect of Ensemble Generalization on the Highest-Occupied {Kohn--Sham} Eigenvalue},
  journal = {Journal of Chemical Physics}, year = {2015}, volume = {143}, pages = {104105},
  doi = {10.1063/1.4930119}}

@article{Kraisler2015PRA,
  author = {Kraisler, Eli and Kronik, Leeor},
  title = {Elimination of the asymptotic fractional dissociation problem in {Kohn--Sham} density-functional theory using the ensemble-generalization approach},
  journal = {Physical Review A}, year = {2015}, volume = {91}, pages = {032504},
  doi = {10.1103/PhysRevA.91.032504}}

@article{Lehtola2020,
  author = {Lehtola, Susi},
  title = {Fully numerical calculations on atoms with fractional occupations and range-separated exchange functionals},
  journal = {Physical Review A}, year = {2020}, volume = {101}, pages = {012516},
  doi = {10.1103/PhysRevA.101.012516}}

@article{LiLuYang2017,
  author = {Li, Chen and Lu, Jianfeng and Yang, Weitao},
  title = {On extending {Kohn-Sham} density functionals to systems with fractional number of electrons},
  journal = {Journal of Chemical Physics}, year = {2017}, volume = {146}, pages = {214109},
  doi = {10.1063/1.4982951}}

@article{LiYang2017,
  author = {Li, Chen and Yang, Weitao},
  title = {On the piecewise convex or concave nature of ground state energy as a function of fractional number of electrons for approximate density functionals},
  journal = {Journal of Chemical Physics}, year = {2017}, volume = {146}, pages = {074107},
  doi = {10.1063/1.4974988}}

@article{Lieb1981,
  author = {Lieb, Elliott H.},
  title = {Variational Principle for Many-Fermion Systems},
  journal = {Physical Review Letters}, year = {1981}, volume = {46}, pages = {457--459},
  doi = {10.1103/PhysRevLett.46.457}}

@book{LinderbergOhrn1973,
  author = {Linderberg, Jan and {\"O}hrn, Yngve},
  title = {Propagators in Quantum Chemistry},
  publisher = {Academic Press}, address = {London}, year = {1973}}

@article{LykosPratt1963,
  author = {Lykos, P. and Pratt, G. W.},
  title = {Discussion on the {Hartree--Fock} Approximation},
  journal = {Reviews of Modern Physics}, year = {1963}, volume = {35}, pages = {496--501},
  doi = {10.1103/RevModPhys.35.496},
  note = {Comment by P.-O. L{\"o}wdin on the symmetry properties of {Hartree--Fock} functions}}

@article{Mermin1963,
  author = {Mermin, N. David},
  title = {Stability of the thermal {Hartree-Fock} approximation},
  journal = {Annals of Physics}, year = {1963}, volume = {21}, pages = {99--121},
  doi = {10.1016/0003-4916(63)90226-4}}

@article{MilgromSegal2002,
  author = {Milgrom, Paul and Segal, Ilya},
  title = {Envelope Theorems for Arbitrary Choice Sets},
  journal = {Econometrica}, year = {2002}, volume = {70}, pages = {583--601},
  doi = {10.1111/1468-0262.00296}}

@article{MoriSanchez2008PRL,
  author = {Mori-S{\'a}nchez, Paula and Cohen, Aron J. and Yang, Weitao},
  title = {Localization and Delocalization Errors in Density Functional Theory and Implications for Band-Gap Prediction},
  journal = {Physical Review Letters}, year = {2008}, volume = {100}, pages = {146401},
  doi = {10.1103/PhysRevLett.100.146401}}

@article{MoriSanchez2009,
  author = {Mori-S{\'a}nchez, Paula and Cohen, Aron J. and Yang, Weitao},
  title = {Discontinuous Nature of the Exchange-Correlation Functional in Strongly Correlated Systems},
  journal = {Physical Review Letters}, year = {2009}, volume = {102}, pages = {066403},
  doi = {10.1103/PhysRevLett.102.066403}}

@article{Noodleman1981,
  author = {Noodleman, Louis},
  title = {Valence bond description of antiferromagnetic coupling in transition metal dimers},
  journal = {Journal of Chemical Physics}, year = {1981}, volume = {74}, pages = {5737--5743},
  doi = {10.1063/1.440939}}

@article{Pastorczak2013,
  author = {Pastorczak, Ewa and Gidopoulos, Nikitas I. and Pernal, Katarzyna},
  title = {Calculation of electronic excited states of molecules using the {Helmholtz} free-energy minimum principle},
  journal = {Physical Review A}, year = {2013}, volume = {87}, pages = {062501},
  doi = {10.1103/PhysRevA.87.062501}}

@article{PastorczakPernal2014,
  author = {Pastorczak, Ewa and Pernal, Katarzyna},
  title = {Ensemble density variational methods with self- and ghost-interaction-corrected functionals},
  journal = {Journal of Chemical Physics}, year = {2014}, volume = {140}, pages = {18A514},
  doi = {10.1063/1.4866998}}

@article{Peach2015,
  author = {Peach, Michael J. G. and Teale, Andrew M. and Helgaker, Trygve and Tozer, David J.},
  title = {Fractional Electron Loss in Approximate {DFT} and {Hartree--Fock} Theory},
  journal = {Journal of Chemical Theory and Computation}, year = {2015}, volume = {11}, pages = {5262--5268},
  doi = {10.1021/acs.jctc.5b00804}}

@article{Perdew1982,
  author = {Perdew, John P. and Parr, Robert G. and Levy, Mel and Balduz, Jose L.},
  title = {Density-Functional Theory for Fractional Particle Number: Derivative Discontinuities of the Energy},
  journal = {Physical Review Letters}, year = {1982}, volume = {49}, pages = {1691--1694},
  doi = {10.1103/PhysRevLett.49.1691}}

@article{PerezMartin2008,
  author = {Perez-Martin, Sara and Robledo, L. M.},
  title = {Microscopic justification of the equal filling approximation},
  journal = {Physical Review C}, year = {2008}, volume = {78}, pages = {014304},
  doi = {10.1103/PhysRevC.78.014304}}

@article{PopleNesbet1954,
  author = {Pople, J. A. and Nesbet, R. K.},
  title = {Self-Consistent Orbitals for Radicals},
  journal = {Journal of Chemical Physics}, year = {1954}, volume = {22}, pages = {571--572},
  doi = {10.1063/1.1740120}}

@article{PySCF2020,
  author = {Sun, Qiming and Zhang, Xing and Banerjee, Samragni and Bao, Peng and Barbry, Marc and Blunt, Nick S. and Bogdanov, Nikolay A. and Booth, George H. and Chen, Jia and Cui, Zhi-Hao and Eriksen, Janus J. and Gao, Yang and Guo, Sheng and Hermann, Jan and Hermes, Matthew R. and Koh, Kevin and Koval, Peter and Lehtola, Susi and Li, Zhendong and Liu, Junzi and Mardirossian, Narbe and McClain, James D. and Motta, Mario and Mussard, Bastien and Pham, Hung Q. and Pulkin, Artem and Purwanto, Wirawan and Robinson, Paul J. and Ronca, Enrico and Sayfutyarova, Elvira R. and Scheurer, Maximilian and Schurkus, Henry F. and Smith, James E. T. and Sun, Chong and Sun, Shi-Ning and Upadhyay, Shiv and Wagner, Lucas K. and Wang, Xiao and White, Alec and Whitfield, James Daniel and Williamson, Mark J. and Wouters, Sebastian and Yang, Jun and Yu, Jason M. and Zhu, Tianyu and Berkelbach, Timothy C. and Sharma, Sandeep and Sokolov, Alexander Yu. and Chan, Garnet Kin-Lic},
  title = {Recent Developments in the {PySCF} Program Package},
  journal = {Journal of Chemical Physics}, year = {2020}, volume = {153}, pages = {024109},
  doi = {10.1063/5.0006074}}

@article{RabuckScuseria1999,
  author = {Rabuck, Angela D. and Scuseria, Gustavo E.},
  title = {Improving self-consistent field convergence by varying occupation numbers},
  journal = {Journal of Chemical Physics}, year = {1999}, volume = {110}, pages = {695--700},
  doi = {10.1063/1.478177}}

@article{Reiher2001,
  author = {Reiher, Markus and Salomon, Oliver and Hess, Bernd Artur},
  title = {Reparameterization of hybrid functionals based on energy differences of states of different multiplicity},
  journal = {Theoretical Chemistry Accounts}, year = {2001}, volume = {107}, pages = {48--55},
  doi = {10.1007/s00214-001-0300-3}}

@article{RobinDay1968,
  author = {Robin, Melvin B. and Day, Peter},
  title = {Mixed Valence Chemistry---A Survey and Classification},
  journal = {Advances in Inorganic Chemistry and Radiochemistry}, year = {1968}, volume = {10}, pages = {247--422},
  doi = {10.1016/S0065-2792(08)60179-X}}

@article{Roothaan1960,
  author = {Roothaan, C. C. J.},
  title = {Self-Consistent Field Theory for Open Shells of Electronic Systems},
  journal = {Reviews of Modern Physics}, year = {1960}, volume = {32}, pages = {179--185},
  doi = {10.1103/RevModPhys.32.179}}

@article{SeegerPople1977,
  author = {Seeger, Rolf and Pople, John A.},
  title = {Self-consistent molecular orbital methods. {XVIII}. Constraints and stability in {Hartree--Fock} theory},
  journal = {Journal of Chemical Physics}, year = {1977}, volume = {66}, pages = {3045--3050},
  doi = {10.1063/1.434318}}

@article{SenjeanFromager2018,
  author = {Senjean, Bruno and Fromager, Emmanuel},
  title = {Unified formulation of fundamental and optical gap problems in density-functional theory for ensembles},
  journal = {Physical Review A}, year = {2018}, volume = {98}, pages = {022513},
  doi = {10.1103/PhysRevA.98.022513}}

@article{Slater1969,
  author = {Slater, J. C. and Mann, J. B. and Wilson, T. M. and Wood, J. H.},
  title = {Nonintegral Occupation Numbers in Transition Atoms in Crystals},
  journal = {Physical Review}, year = {1969}, volume = {184}, pages = {672--694},
  doi = {10.1103/PhysRev.184.672}}

@incollection{Slater1972,
  author = {Slater, J. C.},
  title = {Statistical Exchange-Correlation in the Self-Consistent Field},
  booktitle = {Advances in Quantum Chemistry},
  publisher = {Academic Press}, year = {1972}, volume = {6}, pages = {1--92},
  doi = {10.1016/S0065-3276(08)60541-9}}

@book{SzaboOstlund,
  author = {Szabo, Attila and Ostlund, Neil S.},
  title = {Modern Quantum Chemistry: Introduction to Advanced Electronic Structure Theory},
  publisher = {Dover Publications}, address = {Mineola, NY}, year = {1996}}

@article{Theophilou1979,
  author = {Theophilou, A. K.},
  title = {The energy density functional formalism for excited states},
  journal = {Journal of Physics C: Solid State Physics}, year = {1979}, volume = {12}, pages = {5419--5430},
  doi = {10.1088/0022-3719/12/24/013}}

@article{ThomHeadGordon2008,
  author = {Thom, Alex J. W. and Head-Gordon, Martin},
  title = {Locating Multiple Self-Consistent Field Solutions: An Approach Inspired by Metadynamics},
  journal = {Physical Review Letters}, year = {2008}, volume = {101}, pages = {193001},
  doi = {10.1103/PhysRevLett.101.193001}}

@article{Valone1980,
  author = {Valone, Steven M.},
  title = {Consequences of extending 1-matrix energy functionals from pure-state representable to all ensemble representable 1 matrices},
  journal = {Journal of Chemical Physics}, year = {1980}, volume = {73}, pages = {1344--1349},
  doi = {10.1063/1.440249}}

@article{Vydrov2007,
  author = {Vydrov, Oleg A. and Scuseria, Gustavo E. and Perdew, John P.},
  title = {Tests of functionals for systems with fractional electron number},
  journal = {Journal of Chemical Physics}, year = {2007}, volume = {126}, pages = {154109},
  doi = {10.1063/1.2723119}}

@article{Weigend2005,
  author = {Weigend, Florian and Ahlrichs, Reinhart},
  title = {Balanced basis sets of split valence, triple zeta valence and quadruple zeta valence quality for {H} to {Rn}: Design and assessment of accuracy},
  journal = {Physical Chemistry Chemical Physics}, year = {2005}, volume = {7}, pages = {3297--3305},
  doi = {10.1039/b508541a}}

@article{WernerMeyer1981,
  author = {Werner, Hans-Joachim and Meyer, Wilfried},
  title = {A quadratically convergent {MCSCF} method for the simultaneous optimization of several states},
  journal = {Journal of Chemical Physics}, year = {1981}, volume = {74}, pages = {5794--5801},
  doi = {10.1063/1.440892}}

@article{Yang2000,
  author = {Yang, Weitao and Zhang, Yingkai and Ayers, Paul W.},
  title = {Degenerate Ground States and a Fractional Number of Electrons in Density and Reduced Density Matrix Functional Theory},
  journal = {Physical Review Letters}, year = {2000}, volume = {84}, pages = {5172--5175},
  doi = {10.1103/PhysRevLett.84.5172}}

@article{Yang2013,
  author = {Yang, Weitao and Mori-S{\'a}nchez, Paula and Cohen, Aron J.},
  title = {Extension of many-body theory and approximate density functionals to fractional charges and fractional spins},
  journal = {Journal of Chemical Physics}, year = {2013}, volume = {139}, pages = {104114},
  doi = {10.1063/1.4817183}}

@article{Zerner1989,
  author = {Zerner, Michael C.},
  title = {A configuration-averaged {Hartree--Fock} procedure},
  journal = {International Journal of Quantum Chemistry}, year = {1989}, volume = {35}, pages = {567--575},
  doi = {10.1002/qua.560350409}}

@article{Ziegler1977,
  author = {Ziegler, Tom and Rauk, Arvi and Baerends, Evert J.},
  title = {On the calculation of multiplet energies by the {Hartree--Fock--Slater} method},
  journal = {Theoretica Chimica Acta}, year = {1977}, volume = {43}, pages = {261--271},
  doi = {10.1007/BF00551551}}

@article{eT2020,
  author = {Folkestad, Sarai D. and Kj{\o}nstad, Eirik F. and Myhre, Rolf H. and Andersen, Josefine H. and Balbi, Alice and Coriani, Sonia and Giovannini, Tommaso and Goletto, Linda and Haugland, Tor S. and Hutcheson, Anders and H{\o}yvik, Ida-Marie and Moitra, Torsha and Paul, Alexander C. and Scavino, Marco and Skeidsvoll, Andreas S. and Tveten, {\AA}smund H. and Koch, Henrik},
  title = {{eT} 1.0: An open source electronic structure program with emphasis on coupled cluster and multilevel methods},
  journal = {Journal of Chemical Physics}, year = {2020}, volume = {152}, pages = {184103},
  doi = {10.1063/5.0004713}}

\end{document}